\documentclass[a4paper,fleqn]{cas-sc}

\usepackage[numbers, sort&compress]{natbib}

\usepackage{float}
\usepackage{amsmath} 
\usepackage{amssymb} 
\usepackage{booktabs}
\usepackage{siunitx}
\usepackage[pagewise]{lineno}
\usepackage{xurl}
\usepackage{caption}
\usepackage{pdfpages} 

\def\tsc#1{\csdef{#1}{\textsc{\lowercase{#1}}\xspace}}
\tsc{WGM}
\tsc{QE}
\tsc{EP}
\tsc{PMS}
\tsc{BEC}
\tsc{DE}

\begin{document}
\renewcommand{\lastpage}{\pageref{LastPageMain}} 
\let\WriteBookmarks\relax
\def\floatpagepagefraction{1}
\def\textpagefraction{.001}

\shorttitle{First-order magnetostructural transition in Ni-Mn-Sn powder - A matter of particle and grain size}

\shortauthors{J. Puy et~al.}

\title [mode = title]{First-order magnetostructural transition in Ni-Mn-Sn powder - A matter of particle and grain size}                      


%
\author[1]{Johannes Puy}[orcid= 0009-0007-6557-5189]
\credit{Writing – original draft, Writing – review \& editing, Visualization, Validation, Methodology, Investigation, Formal analysis, Conceptualization}
\affiliation[1]{organization={Functional Materials, Institute of Materials Science, Technical University of Darmstadt},
    addressline={Peter-Grünberg-Str. 16}, 
    city={Darmstadt},
    postcode={64287}, 
    country={Germany}}

\author[2]{David Koch}[orcid= 0000-0002-6841-9962]
\credit{Writing – review \& editing, Investigation, Formal analysis}
\affiliation[2]{organization={Strukturforschung, Institute of Materials Science, Technical University of Darmstadt},
    addressline={Peter-Grünberg-Str. 4}, 
    city={Darmstadt},
    postcode={64287}, 
    country={Germany}}

\author[3]{Enrico Bruder}[orcid= 0000-0001-7981-0871]
\credit{Writing – review \& editing, Investigation, Formal analysis}
\affiliation[3]{organization={Physikalische~Metallkunde,~Institute~of~Materials~Science,~Technical~University~of~Darmstadt},
    addressline={Peter-Grünberg-Str.~4}, 
    city={Darmstadt},
    postcode={64287}, 
    country={Germany}}

\author[1]{Sagar Ghorai}[orcid= 0000-0003-2790-116X]
\credit{Investigation, Formal analysis}

\author[1]{Benedikt Beckmann}[orcid= 0000-0002-2232-1804]
\credit{Writing – review \& editing, Investigation, Formal analysis}

\author[4]{Heiner Gutte}
\credit{Writing – review \& editing, Resources}
\affiliation[4]{organization={Institute of Iron and Steel Technology, TU Bergakademie Freiberg},
    addressline={Leipziger Str. 34}, 
    city={Freiberg},
    postcode={09599}, 
    country={Germany}}

\author[5]{Volodymyr Baran}[orcid= 0000-0003-2379-3632]
\credit{Writing – review \& editing, Investigation}
\affiliation[5]{organization={Deutsches Elektronen-Synchrotron DESY},
    addressline={Notkestr. 85}, 
    city={Hamburg},
    postcode={22609}, 
    country={Germany}}

\author[1]{Oliver Gutfleisch}[orcid= 0000-0001-8021-3839]
\credit{Writing – review \& editing, Supervision, Resources}
\fnmark[1]
\fntext[fn1]{Oliver Gutfleisch was an editor of the journal during the review period of the article. To avoid a conflict of interest, Oliver Gutfleisch was blinded to the record and another editor processed this manuscript.}

\author[1]{Franziska Scheibel}[orcid= 0000-0001-7981-0871]
\credit{Writing – review \& editing, Writing – original draft, Visualization, Validation, Methodology, Formal analysis, Conceptualization, Supervision, Resources, Project administration, Funding acquisition}
\cormark[1]
\cortext[cor1]{Corresponding author: franziska.scheibel@tu-darmstadt.de}

\begin{abstract}
In Ni-Mn-based Heusler alloys, the magnetocaloric effect originates from a first-order magnetostructural transition (FOMST). However, the influence of particle and grain size on thermal hysteresis, transition width, and nucleation and growth mechanism remains poorly understood, as mechanical grinding can introduce residual stresses and processing-induced defects. In this study, we examined the size dependence of the FOMST in gas-atomized Ni–Mn–Sn powder by combining microstructural characterization with temperature-dependent XRD, magnetometry, calorimetry and single-particle measurements, enabling decoupling of size dependence from processing-related influences. We observed a pronounced size dependence of the transition ranges ($M_s - M_f$) and ($A_f - A_s$), which decrease with decreasing particle size and approach zero for single-crystalline particles below $\varnothing \approx 20-30$\,µm. The thermal hysteresis $\Delta T_{hyst}$  decreases with increasing particle size, reaching a lower limit of 17 \,K for large polycrystalline particles above 100 \,µm. EBSD reveals a nearly constant grain size, implying a transition from polycrystalline to single-crystalline particles as particle size decreases. In polycrystalline particles, multiple grains transform gradually over an extended temperature interval because grain boundaries act as barriers to the progression of transformation. In contrast, single-grain particles lack preferred nucleation sites, resulting in larger hysteresis, while the absence of barriers to transformation propagation causes the entire particle to transform in a narrow range. These results demonstrate that the size dependence of the FOMST is governed by the interplay between particle size, grain boundaries, and the free particle surface.
\end{abstract}




\maketitle
\section{Introduction}
The past decade has seen record-high global mean temperatures, with 2024 being the warmest year on record \cite{WorldMeteorologicalOrganization.2024}. Continued global warming is expected to increase the demand for cooling over the coming decades. Currently used conventional cooling devices rely on the vapor-compression technique, which is more than a hundred years old, and require the use of environmentally harmful refrigerants. A significant increase in cooling efficiency and the elimination of harmful refrigerants demand the search for alternative cooling technologies. Solid-state caloric cooling, utilizing caloric effects, is seen as a promising alternative for the vapor-compression technique \cite{Pecharsky.1997, Glanz.1998, Gutfleisch.2011, Franco.2012, Gutfleisch.2016, Scheibel.2018, Gottschall.2017}. Magnetocaloric cooling uses magnetic solid-state material as refrigerant and operates entirely without refrigerant gases, which makes it a potentially more environmentally friendly and energy-efficient cooling alternative \cite{HWANGY..2022}. \\
Ni-Mn-X (X= In, Sn, Sb, Ti) Heusler alloys are considered as potential caloric refrigerants, as the alloys show an inverse magnetocaloric effect (MCE) and a conventional elastocaloric effect (ECE) making these materials suitable for magnetocaloric, elastocaloric or multicaloric cooling \cite{Krenke.2005b, Planes.2009, Liu.2012, Gottschall.2019, SternTaulats.2018, Beckmann.2025}. The caloric effects arise from the first-order magnetostructural transition (FOMST), involving a martensitic transformation that is driven by a nucleation and growth mechanism \cite{Krenke.2005b, Liu.2012}. The nucleation and growth mechanism and the related FOMST properties, thermal hysteresis $\Delta T_{hyst}$ and transition widths $(M_s - M_f)$ and $(A_f - A_s)$, depend on a variety of intrinsic and extrinsic properties \cite{Gutfleisch.2016, Scheibel.2018}. Chemical composition \cite{Taubel.2018}, transition temperature $T_t$ \cite{Beckmann.2023} and microstructural aspects like homogeneity \cite{Taubel.2020}, grain size \cite{Pfeuffer.2021}, secondary phases \cite{Pfeuffer.2021b, Scheibel.2023b} and texture \cite{Pfeuffer.2021} have a significant impact on the FOMST properties and the related MCE in Ni-Mn-based Heusler alloys. \\
In view of the fact that the active magnetocaloric material is typically incorporated into the refrigeration device in packed powder beds, parallel plate arrays or microchannel geometries with sizes below 500 µm \cite{Waske.2018b}, it is necessary to establish a better understanding of how decreasing the characteristic length scales in terms of sample dimension or particle size impacts the nucleation and growth process of the martensitic transformation. The possibility to process spherical powders by gas-atomization for additive manufacturing techniques like laser powder bed fusion (PBF-LB) or direct energy deposition (DED) \cite{Scheibel.2023, Sun.2022} opened up a new possibility to study the size dependence of the FOMST, and tailor the FOMST properties by selecting a specific particle size. \\
Early studies of size effects on the martensitic transformation in shape memory alloys focused on non-magnetic Cu-based microwires, considering both stress-induced and temperature-induced transformations. Reported effects include changes in transformation temperatures, transformation stresses, hysteresis, and stored elastic energy \cite{Chen.2011, Ueland.2012, Ueland.2013, Tuncer.2015}. Chen \textit{et al.} observed larger hysteresis and greater energy dissipation in thinner wires, which they attributed to enhanced internal frictional work arising from stronger surface pinning of the phase boundary at higher surface-to-volume ratios \cite{Chen.2011}. Shifts in transformation temperatures and stresses, in contrast, were attributed to the relief of elastic transformation stresses at free surfaces \cite{Chen.2011, Ueland.2012}.
Wire diameter has also been reported to affect transformation morphology. Ueland \textit{et al.} observed a transition from multi-domain to single-domain transformation behavior with decreasing wire diameter. In thinner wires, transformation proceeded through the nucleation and subsequent growth of a single martensite plate, whereas thicker wires exhibited continued nucleation and growth of additional plates \cite{Ueland.2013}. \\
Studies regarding the size dependence of the martensitic transformation in Ni-Mn-based Heusler alloys were conducted on thin films \cite{Zhang.2019}, milled particles \cite{Gottschall.2017, Cavazzini.2022, Khanna.2021} and gas-atomized powders \cite{Zhang.2021, Lago.2024}. For Ni-Mn-Sn thin films, Zhang \textit{et al.} found that $\Delta T_{hyst}$ of the films does not depend on the grain size, but increases with decreasing film thickness according to a power law model \cite{Zhang.2019}. In another study on gas-atomized Ni-Co-Mn-Sn single particles (38 µm to 9 µm), Zhang \textit{et al.} confirm the trend of an increasing $\Delta T_{hyst}$ with decreasing particle size \cite{Zhang.2021}, which is also found by Lago \textit{el al.} in gas-atomized Ni-Co-Mn-In single particles (42 µm to 8 µm) \cite{Lago.2024}. In each case, the authors attribute this to greater energy dissipation through enhanced frictional work in smaller particles and a limitation of nucleation due to a reduced probability of finding potent defects acting as nucleation sites \cite{Zhang.2021, Lago.2024}. Additionally, different types of transformation characteristics are revealed in these particles, ranging from continuous transformations with broader transition temperature ranges ($>$15 K) to abrupt transformations with narrow ranges ($<$1 K). \\
In addition to the studies on atomized spherical powder, several studies on milled Ni-Mn-based Heusler alloy particles have been reported, although grinding and milling can introduce plastic deformation and lattice defects, such as dislocations, and generate internal stresses and associated lattice strains in these alloys \cite{Gottschall.2017, Cavazzini.2022, Khanna.2021, LopezGarcia.2018, Lopez-Garcia2021EffectHighenergyBallmilling, SanchezAlarcos.2019, Unzueta2017119SnMossbauerSpectroscopy, Tian2008EffectBallMilling}. For irregular large particles, Gottschall \textit{et al.} describe a significant broadening of $\Delta T_{hyst}$ and $(M_s - M_f)$ and $(A_f - A_s)$ with decreasing particle size (250 µm to 100 µm) \cite{Gottschall.2017}. Cavazzini \textit{et al.} demonstrate a size dependence of the martensitic transformation in hand-ground and annealed Ni-Mn-In-Co particles (250 µm to 66 µm) \cite{Cavazzini.2022}. Specifically, $\Delta T_{hyst}$ is significantly broadened as a result of a broadening of $(M_s - M_f)$ towards lower temperatures, whereas $M_s$ and $(A_f - A_s)$ are not affected by particle size. In addition, an increased amount of retained austenite in smaller particles below $M_f$ is confirmed by magnetometry and XRD, which the authors attribute to larger residual stress in smaller particles, as a direct result of the powder-synthesis process. As the particle size decreases, these residual stresses increasingly suppress the martensitic transformation and thus lead to a broadening of the hysteresis. In another recent study, Khanna \textit{et al.} also describe a correlation between particle size and degree of deformation in milled Ni-Mn-In-Co particles, which leads to a broadening of the FOMST \cite{Khanna.2021}. \\
Besides lattice defects like dislocations and internal stress and strain, the defects introduced by mechanical processing can also include antiphase boundaries associated with dislocations. In ball-milled Ni-Mn-Ga, the dissociation of full dislocations into partial dislocations and their subsequent motion have been proposed to produce antiphase boundaries or reduce atomic order, depending on the displacement vector \cite{Tian2008EffectBallMilling}. Such defects can affect magnetic properties even without detectable atomic disordering. In mechanically treated Ni-Mn-Sn, a reduction in saturation magnetization was attributed to antiferromagnetic coupling between Mn-atoms across dislocation-associated antiphase boundaries \cite{Unzueta2017119SnMossbauerSpectroscopy}. Under sufficiently severe milling conditions, atomic disordering can additionally contribute to the suppression of the martensitic transformation \cite{Tian2008EffectBallMilling, Cavazzini.2022}. Substantial lattice distortions can arise even during hand grinding. For example, a microstrain of up to 0.93 \% has been reported in hand-ground Ni-Mn-Sn particles \cite{LopezGarcia.2018}. The extent of these processing-induced changes depend on the milling conditions \cite{Cavazzini.2022}. These findings illustrate the difficulty of disentangling particle size effects from the effects of lattice defects, internal stresses and strains, and particle morphology. \\
To disentangle the influence of particle size from effects related to powder morphology and processing-induced defects, such as dislocations, associated antiphase boundaries or atomic disorder \cite{Tian2008EffectBallMilling, Cavazzini.2022, LopezGarcia.2018, Unzueta2017119SnMossbauerSpectroscopy}, we used gas-atomized Ni-Mn-Sn powder in this study. Owing to its spherical morphology and preparation without mechanical grinding or milling, this material system provides a suitable model to systematically investigate the size dependence of the FOMST while avoiding the additional defects and internal stresses introduced by these processing steps. Despite previous studies, it remains unclear how particle size and the associated grain structure influence the FOMST, particularly with respect to thermal hysteresis, transition widths, and the underlying nucleation and growth mechanisms. Zhang \textit{et al.} \cite{Zhang.2021} investigated individual Ni-Co-Mn-Sn particles, including larger polycrystalline particles, but their statistical analysis of transformation behaviour focused only on particles classified as single crystals, with diameters of approximately 9–38\,µm. The present study examines individual particles over a wider diameter range of 13–127\,µm, including both single-crystalline and polycrystalline particles. We relate their transformation behaviour to particle microstructure and direct in-situ observations of the martensite nucleation and growth. In this context, we hypothesize that the observed size dependence of the FOMST is closely linked to the microstructure and presence of grain boundaries within a particle, which is expected to affect the nucleation and growth behavior of the martensitic transformation.
To test this hypothesis, we conducted a systematic study of gas-atomized Ni-Mn-Sn powders. The FOMST is characterized by a combination of structural (XRD), microstructural (SEM, EBSD), calorimetric (DSC), and magnetic measurements, including both powder-based and single-particle analyses. Within this framework, we identified a correlation between particle size and crystallinity, providing the basis for a detailed investigation of size-dependent transformation behavior for $\Delta T_{hyst}$ and $(M_s - M_f)$ and $(A_f - A_s)$.

\section{Experimental Methods}

The spherical powder was processed by Vacuum inert gas atomization (VIGA) at the TU Freiberg using Ni\textsubscript{50}Mn\textsubscript{50} pre-alloy and elemental Ni and Sn pellets with a purity of 99.9\,\%, 99.99\,\% and 99.9\,\%, respectively. For a nominal Ni\textsubscript{50}Mn\textsubscript{36}Sn\textsubscript{14} composition 0.98\,kg Ni, 4.30\,kg Ni\textsubscript{50}Mn\textsubscript{50} and 1.72\,kg Sn are alloyed for three hours. Afterward, the molten metal was gas-atomized. More information on the VIGA device and process can be found in \cite{perminov_manufacturing_2021}. The as-atomized powder exhibits a spherical shape with a small amount of satellites and a Gaussian-type size-distribution with a D\textsubscript{50} value of 33\,µm. 10\,\% of the powder has a smaller diameter then 10\,µm (D\textsubscript{10}) and 90\,\% of the powder has a smaller diameter than 92\,µm (D\textsubscript{90}); more information on the as-atomized powder can be found in \cite{Scheibel.2023}. For the current study the powder was sieved into five different size fractions: 150-75\,µm, 75-63\,µm, 63-50\,µm, 50-20\,µm and $<$20\,µm using cloths with mesh sizes of 150\,µm, 75\,µm, 63\,µm, 50\,µm and 20\,µm. The powder was sieved using a compact screening machine \textit{KSM 500} from \textit{Assonic}. Each size fraction was sieved twice to minimize the residual content of fine particles. The powders of all size fractions were subsequently homogenized by a heat treatment. The powders were wrapped in steel foil, encapsulated in quartz-glass ampules and annealed at 1173\,K for 6\,h under Argon atmosphere followed by quenching in water.\par

The particle size distribution of the different size fractions was determined by laser diffraction using a \textit{Mastersize 3000} from \textit{Malvern Instruments Ltd}. The diameter of the individual particles for the magnetic characterization was determined by optical microscopy. Backscattered electron (BSE) images of the different size fractions were obtained using a \textit{Tescan Vega3} scanning electron microscope (SEM). The grain size distribution and texture were evaluated by electron backscatter diffraction (EBSD) using a FEG-SEM (\textit{Tescan Mira3}) with an additional four-quadrant BSE (4Q-BSE) detector for enhanced microstructural feature analysis through electron channeling contrast imaging (ECCI). The EBSD measurements were performed with a step size of 500\,nm and 250\,nm for particle sizes above and below 20\,µm, respectively, and 8\,ms acquisition time per step. For this analysis, the loose powder was mixed with conductive epoxy \textit{Technovit5000} and mechanically ground down to 5\,µm grit size. After grinding, the samples were vibration polished for 6\,h with a 0.02\,µm colloidal SiO\textsubscript{2} polishing suspension. The chemical composition of each size fraction was determined by inductively coupled plasma optical emission spectroscopy (ICP-OES). Approximately 100\,mg of powder was dissolved in 2\,ml H\textsubscript{2}O + 3\,ml 65\,\% HNO\textsubscript{3} + 1\,ml 32\,\% HCL. ICP-OES was performed using an \textit{iCAP PRO ICP-OES} system from \textit{ThermoFisher} and standard solutions containing 200, 150, 100, and 50\,ppm of Ni, Mn, and Sn. Temperature-dependent X-ray diffraction (XRD) for structural characterization was performed at the P02.1 Petra III Beamline at DESY using a synchrotron beam radiation of 0.207\,\AA  \cite{Dippel.2015}. The data were collected by a \textit{Varex XRD 4343CT} detector (150 × 150\,µm\textsuperscript{2} pixel size, 2880 x 2880 pixel area) using a detector distance of 1700\,mm and an exposure time of 45\,s. Two absorbers were used due to strong scattering. The measurements were performed in the $2\theta$-range of 0.0239$^{\circ}$-15.7431$^{\circ}$ and a temperature range between 300\,K and 150\,K. Temperature steps of 3\,K were used in the transition region between 230\,K and 285\,K, and 15\,K outside this region. The data were collected during cooling and heating conditions. The Rietveld refinement was done with \textit{JANA2006} \cite{petricek_crystallographic_2014} applying the superspace approach \cite{righi_crystal_2008}.\par

Temperature-dependent vibrating sample magnetometry (VSM) measurements were performed to determine the magnetization and transition temperatures of the powders and individual particles in a temperature range between 150\,K and 350\,K in a constant field of 1\,T using a rate of 2\,Kmin\textsuperscript{-1}. Field-dependent VSM measurements were performed above (300\,K) and below (200\,K) the FOMST in a field range of \(\pm\)5\,T. The measurements were performed using a \textit{Magnetic Property Measurement System (MPMS)} from \textit{Quantum Design} with a maximum field of 7\,T. For the powder analysis, approximately 25\,mg of loose powder was filled into a capsule and inserted into a brass sample holder. For the individual measurements of single particles, individual particles were picked up with a toothpick dipped in \textit{VGE-7031 varnish} and were transferred to a \textit{Kapton} tape. The tape with the particle was then mounted on a quartz half-rod sample holder. The thermal hysteresis $\Delta T_{hyst}$ was determined by the difference of the maxima of the first derivative $dM(T)/dT$ of the temperature-dependent magnetization curve. The Curie temperature $T_c^A$ of the austenite corresponds to the minimum in this curve. The critical temperatures of the FOMST ($A_s$, $A_f$, $M_s$ and $M_f$) were determined by the tangent method \cite{Taubel.2018}. The mean transition temperature $T_t$ was defined as the midpoint of the thermal hysteresis. \par 

The differential scanning calorimetry (DSC) measurements were performed using a silver furnace in a \textit{Netzsch DSC 404 F1 Pegasus}, equipped with a liquid nitrogen cooling system. The measurements were performed in a temperature range between 170\,K and 370\,K, and DSC curves were determined under heating and cooling conditions using a rate of 5\,Kmin\textsuperscript{-1}. For the measurement, a thin layer of powder was distributed in an Al crucible. The transition entropy change $\Delta s_t$ was determined for the forward and the reverse martensitic transformation by integration of the mass-specific heat flow peak originating from the phase transition:

    \begin{equation}
        \Delta s_t = \int (\dot{Q} - \dot{Q}_{baseline}) \cdot \Bigg( \frac{dT}{dt} \Bigg)^{-1} dT
        \label{eq:dS-DSC}
    \end{equation}

\section{Results}
\subsection{Morphology, microstructure and chemical composition}
\begin{figure}[pos=H]
	\centering
		\includegraphics[width=\textwidth]{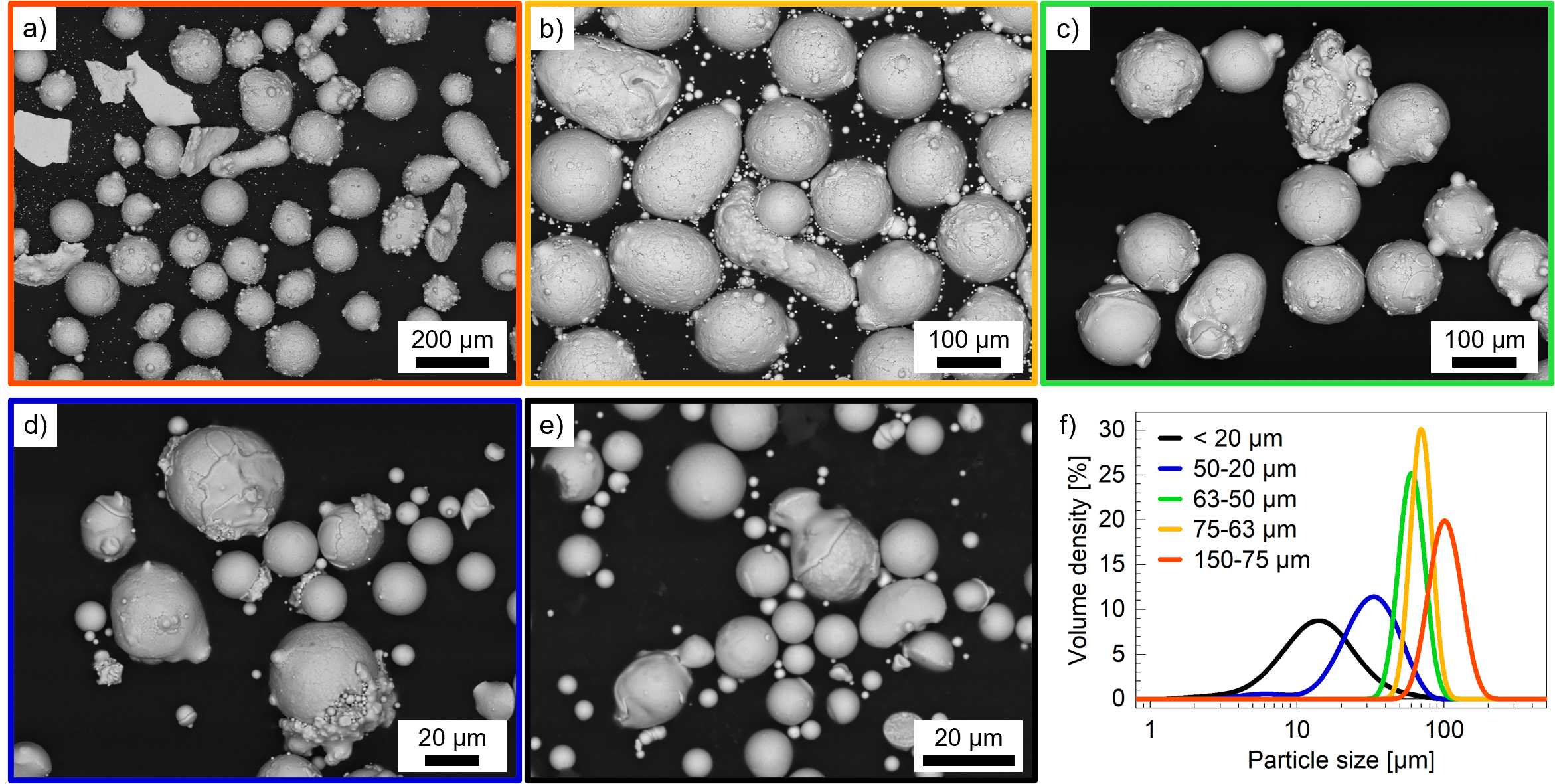}
	\caption{BSE images of the gas-atomized Ni-Mn-Sn powder with particle sizes of a) 150-75\,µm, b) 75-63\,µm, c) 63-50\,µm, d) 50-20\,µm, e) $<$20\,µm. f) Size distribution of the powder with the corresponding particle sizes.}
	  \label{fig:SEM_PartSize}
\end{figure}

\begin{figure}[pos=H]
	\centering
		\includegraphics[width=\textwidth]{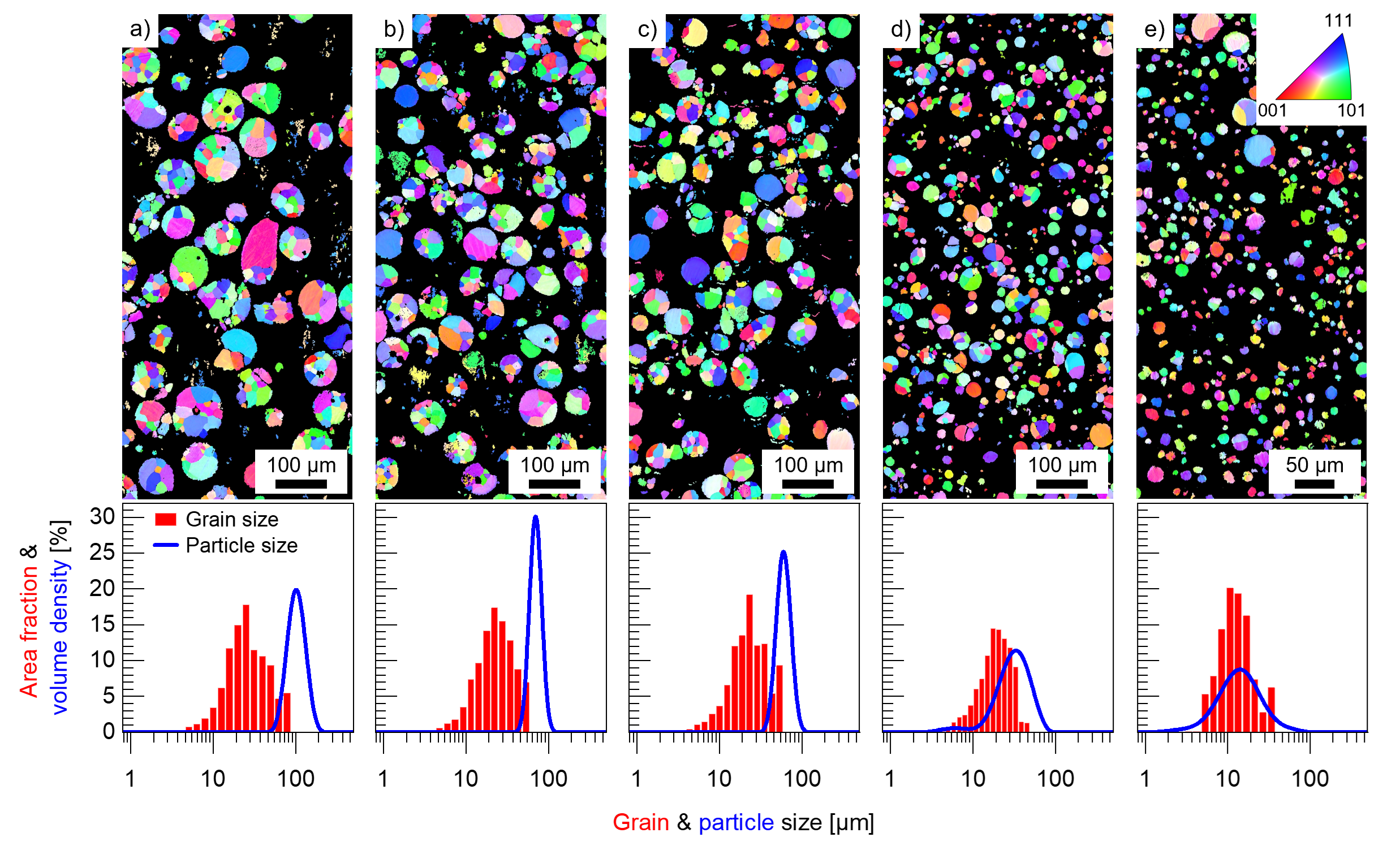}
  \caption{EBSD inverse pole figure maps of gas-atomized Ni-Mn-Sn and the respective grain size and particle size distribution of the powders: a) 150-75\,µm, b) 75-63\,µm, c) 63-50\,µm, d) 50-20\,µm, e) $<$20\,µm. The standard crystallographic triangle is given in e) to relate grain orientation color to the vector normal of the crystallographic plane.}
  \label{fig:EBSD_GrainSize}
\end{figure}

Particle size and shape, chemical composition, and microstructure affect the transition temperatures, transition widths, and thermal hysteresis. The following section presents the characterization and comparison of properties across particle sizes. Figure \ref{fig:SEM_PartSize} a-e) shows the BSE images of the heat-treated powder with the size fractions 150-75\,µm, 75-63\,µm, 63-50\,µm, 50-20\,µm, and $<$20\,µm, as well as the size distribution of the powders. For all powders, spherical particles can be observed. However, smaller amounts of splat cap, elongated, and agglomerated particles can also be observed \cite{williams2017}. Only for the powder 150-75\,µm are also flake-like particles observed. For particles above 20\,µm, besides the spherical-shaped particles, the second most observed shape is satellite particles. These particles are formed after the collisions of a large and a small particle \cite{wang2019a}. All powders exhibit a Gaussian-type particle size distribution, shown in Figure \ref{fig:SEM_PartSize} f). For the powder 50-20\,µm, a minor second peak at approximately 10\,µm can be observed. The presence of the smaller particles can be explained by an agglomeration of very fine particles or by high cohesiveness, where fine particles are attached to larger particles. The median D\textsubscript{50}, D\textsubscript{10} and D\textsubscript{90} values are shown in Table S1 in the supplementary information.
To determine the grain size as a function of particle size, EBSD analyses are performed on the particle cross-sections. Figure \ref{fig:EBSD_GrainSize} shows the inverse pole figure (IPF) maps of the cross sections of the heat-treated particles for the different powders 150-75\,µm, 75-63\,µm, 63-50\,µm, 50-20\,µm and $<$20\,µm. The color coding standard triangle is shown in Figure \ref{fig:EBSD_GrainSize} e), indexing the grain orientation for the austenite phase (Fm-3m crystal structure with a lattice parameter of 0.5986\,nm). The IPF maps indicate that the majority of particles between 150\,µm and 30\,µm contain multiple grains (2-20) with arbitrary orientation. In contrast, the majority of the particles with a size below 30\,µm are single-crystalline. This is confirmed by the comparison between the grain size (blue) and particle size (red) distribution shown in Figure \ref{fig:EBSD_GrainSize} below the IPF maps of the corresponding powders. The particle size and grain size distribution coincide for the powder with a particle diameter below 20\,µm (Figure \ref{fig:EBSD_GrainSize} e)). This confirms that the majority of the particles contain one grain and only a minority contain two or more grains. For the powders with a particle size above 20\,µm, the grain size distributions are comparable, with an average grain size between 13.4 and 19.0\,µm. The cross-sections of the particles also reveal that some particles contain pores. The pores are indicated by the black spherical area within the particles, and are mainly found in the size fractions 150-75\,µm and 75-63\,µm (Figure \ref{fig:EBSD_GrainSize} a) and b)). The cracks along the grain boundaries, which can be observed in some of the larger particles, are caused by the sample preparation.
The chemical composition of the different powder size fractions was determined by ICP-OES, Table \ref{tab:icp} shows the Ni-, Mn- and Sn-content for each size fraction. The chemical composition of Ni\textsubscript{49.6}Mn\textsubscript{36.9}Sn\textsubscript{13.5} is comparable for all size fractions, and deviations between different size fractions are within the error bar. To confirm the chemical homogeneity of individual particles, EDX analysis was performed on multiple particle cross sections, revealing a homogeneous chemical composition across the entire particle diameter. An example is given in Figure S1 in the Supplementary Information .
Both ICP-OES and EDX confirm that the powders are homogeneous and have the same chemical composition in all size fractions. 
\begin{table}[pos=H]
    \centering
    \caption{Chemical composition of each powder size fraction determined by ICP-OES.}
\label{tab:icp}
\begin{tabular}{c 
S[table-format=2.2 \pm 1.2] 
S[table-format=2.2 \pm 1.2] 
S[table-format=2.2 \pm 1.2]}
\toprule
Particle size [\si{\micro\meter}] & {Ni [at.\%]} & {Mn [at.\%]} & {Sn [at.\%]} \\
\midrule
150--75  & 49.59 \pm 0.14 & 36.97 \pm 0.30 & 13.44 \pm 0.15 \\
75--63   & 49.63 \pm 0.66 & 36.92 \pm 0.63 & 13.45 \pm 0.64 \\
63--50   & 49.64 \pm 0.73 & 36.90 \pm 0.62 & 13.46 \pm 0.74 \\
50--20   & 49.65 \pm 0.24 & 36.90 \pm 0.23 & 13.45 \pm 0.26 \\
$<$20    & 49.63 \pm 0.68 & 36.91 \pm 0.63 & 13.46 \pm 0.79 \\
\bottomrule
\end{tabular}
\end{table}

\subsection{First-order magnetostructural transition}
The magnetostructural transition from low-temperature martensite to high-temperature austenite depends on both intrinsic and extrinsic factors \cite{Gutfleisch.2016, Scheibel.2018}. The following sections present the structural, calorimetric and magnetic characterization of the phase transformation, with particular emphasis on correlations among microstructure, particle size, and FOMST properties, including transition temperatures, thermal hysteresis, and transformation width. In addition, the entropy change associated with the FOMST is determined by DSC.

\subsubsection{Structural characterization}
\begin{figure}[pos=H]
	\centering
		\includegraphics[width=\textwidth]{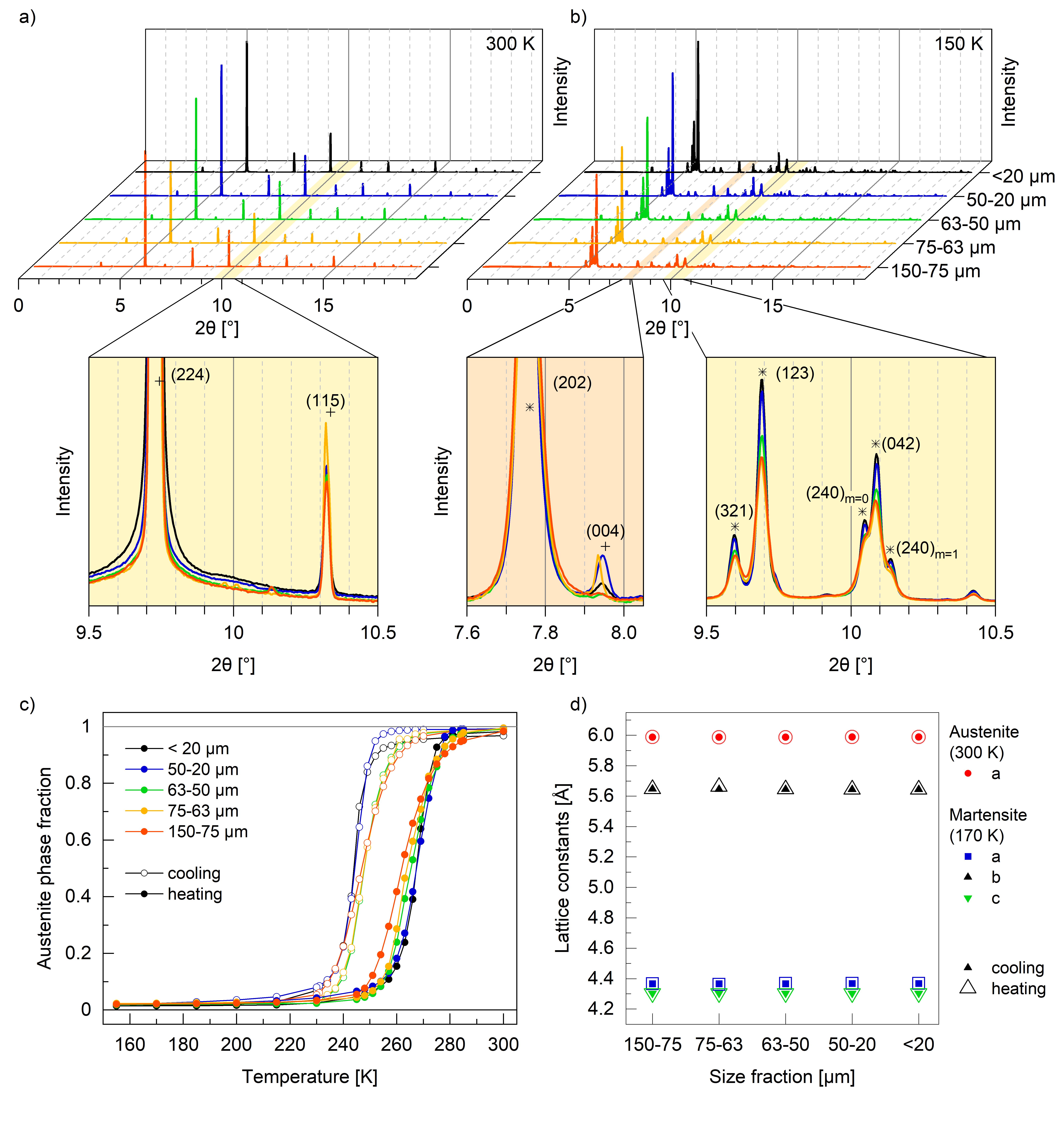}
    \caption{XRD pattern of gas atomized Ni-Mn-Sn at a) 300\,K in L2\textsubscript{1}-austenite phase and at b) 150\,K in 4O-martensite phase. Insets below a) show a broadening of the (224)-reflection with decreasing particle size; insets below b) show (004)-reflection, indicating a minor amount of residual austenite (left) and peak broadening of multiple reflections for larger size fractions (right). Temperature-dependent phase fraction analysis c) reveals a size dependence of the  transformation. Lattice parameters d) of austenite determined at 300\,K and of 4O-martensite determined at 170\,K are independent of size.}
    \label{fig:XRD}
\end{figure}

For the structural characterization of the FOMST, temperature-dependent XRD measurements are performed on the powders over the temperature range from 300\,K to 150\,K under both cooling and heating protocols. Figure \ref{fig:XRD} shows the temperature-dependent XRD patterns of all particle size fractions for a) 300\,K and b) 150\,K. The corresponding lattice parameters of the high-temperature L2\textsubscript{1}-austenite phase and low-temperature 4O-martensite phase for the different powders with particle sizes 150-75\,µm, 75-63\,µm, 63-50\,µm, 50-20\,µm, $<$20\,µm are shown in d) for cooling (full symbol) and heating (open symbol). At 300\,K, all powders are in the high-temperature L2\textsubscript{1}-austenite phase, the XRD pattern shows no indication of a residual martensite phase. The lattice parameter $a$ does not depend on the particle size; Figure \ref{fig:XRD} d) shows a comparable lattice parameter $a$ of 5.99\,\AA\,for all powders before cooling and after heating.\\
A closer view of the XRD pattern for the austenite phase shows a peak broadening of the (224)-reflection (right inset of Figure \ref{fig:XRD} a)) for smaller particle sizes. This broadening can indicate internal stresses. Broadening due to reduced crystallite size can be excluded, as the EBSD analysis shows an average grain size around 15\,µm. For the (224)- and (115)-reflections of L2\textsubscript{1}-austenite, a variation in intensity is observed for the powder 75-63\,µm. Since this powder exhibits the same lattice parameter and the transition behavior is consistent with that of the other powders, we assume that this sample has an uneven distribution of grain orientations that could not be compensated by sample rotation.\\
The diffraction pattern at 150\,K is shown in Figure \ref{fig:XRD} b), corresponding to a 4O-martensite phase with unit cell parameters of $a$\,=\,4.36\,\AA, $b$\,=\,5.64\,\AA\, and $c$\,=\,4.30\,\AA, which is in agreement with values reported in literature \cite{Lin.2016}. At this temperature, only a small amount of residual austenite is detected, as seen in the (004)-reflection at approximately 7.95°, magnified in the middle inset. The highest fractions of residual austenite are observed in the 75-63\,µm and 50-20\,µm size fractions, but no consistent particle-size dependence is evident. The right inset shows the 4O-martensite reflections in the range 9.5-10.5°, highlighting peak broadening with increasing particle size. For the (321)- and (123)-reflections, as well as for the (240)- and (042)-reflection, an increase in peak intensity and sharpness proportional to the decrease of particle size is observable. This can be clearly seen in the left and right shoulders of the (042)-reflection, where the peak overlaps with the (240)-reflection (modulation m\,=\,0) and the first-order satellite of the (240)-reflection (modulation m\,=\,1), and peak broadening leads to peak congestion at larger particle sizes. As with the austenite phase, peak broadening can indicate internal stresses. However, for the 4O-martensite phase, the trend is directly opposite. The particles with the smallest size exhibit the least stress, suggesting reduced stress accumulation during the martensitic transformation, potentially due to more effective accommodation of transformation strains at the free particle surface and reduced internal constraints.\\
The temperature-dependent phase fraction of austenite and martensite is shown in Figure \ref{fig:XRD} c). At 300\,K, all powders exhibit the L2\textsubscript{1}-austenite phase. For temperatures below 200\,K, all powders fully transform into 4O-martensite. The phase fraction analysis reveals differences in transition temperature ranges and thermal hysteresis among the different size fractions. With increasing particle size, $(M_s - M_f)$ and $(A_f - A_s)$ increase from 11\,K for the size fraction $<$20\,µm to 23\,K for the size fraction 150–75\,µm, while $\Delta T_{hyst}$ decreases from 23\,K to 15\,K, respectively.

\subsubsection{Magnetometry}
\begin{figure}
	\centering
		\includegraphics[width=\textwidth]{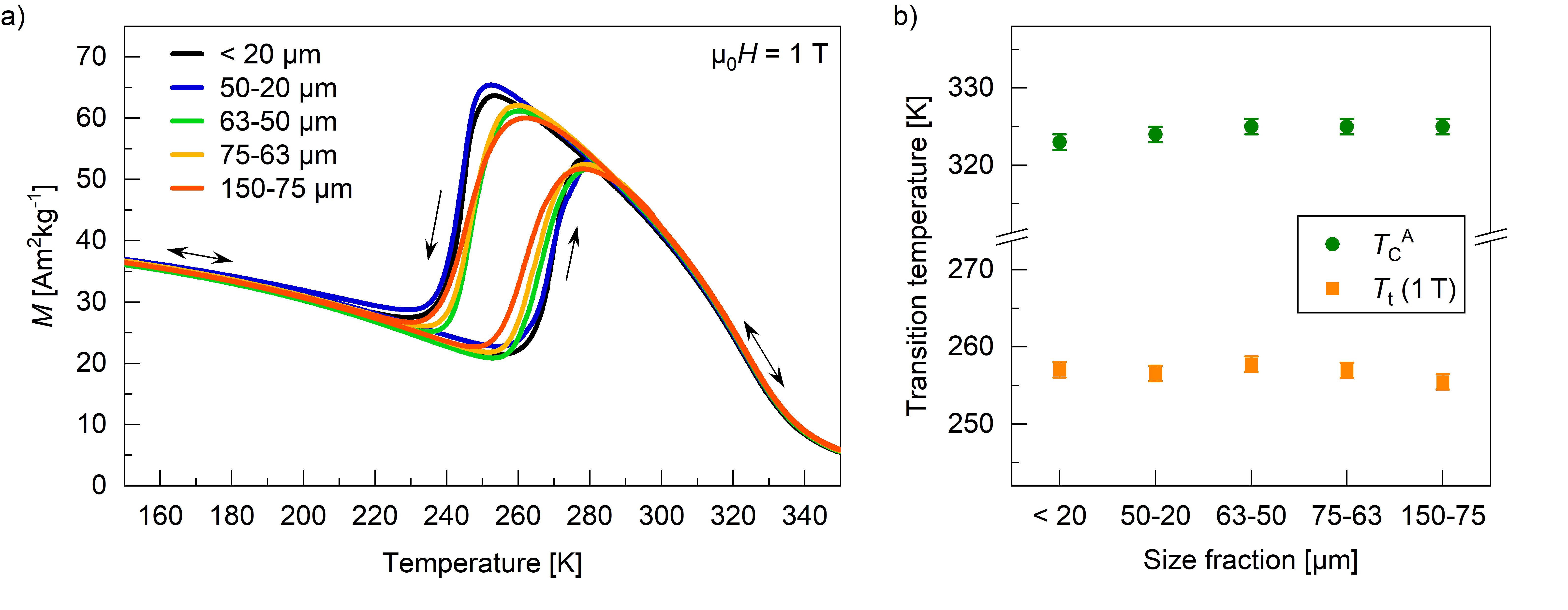}
    \caption{a) Magnetization curves of all powder size fractions as a function of temperature under a magnetic field of 1\,T. b) Curie temperature $T_c^A$ and transition temperature $T_t$ show no dependence on particle size.}
    \label{fig:MT}
\end{figure}

\begin{figure}
	\centering
		\includegraphics[width=\textwidth]{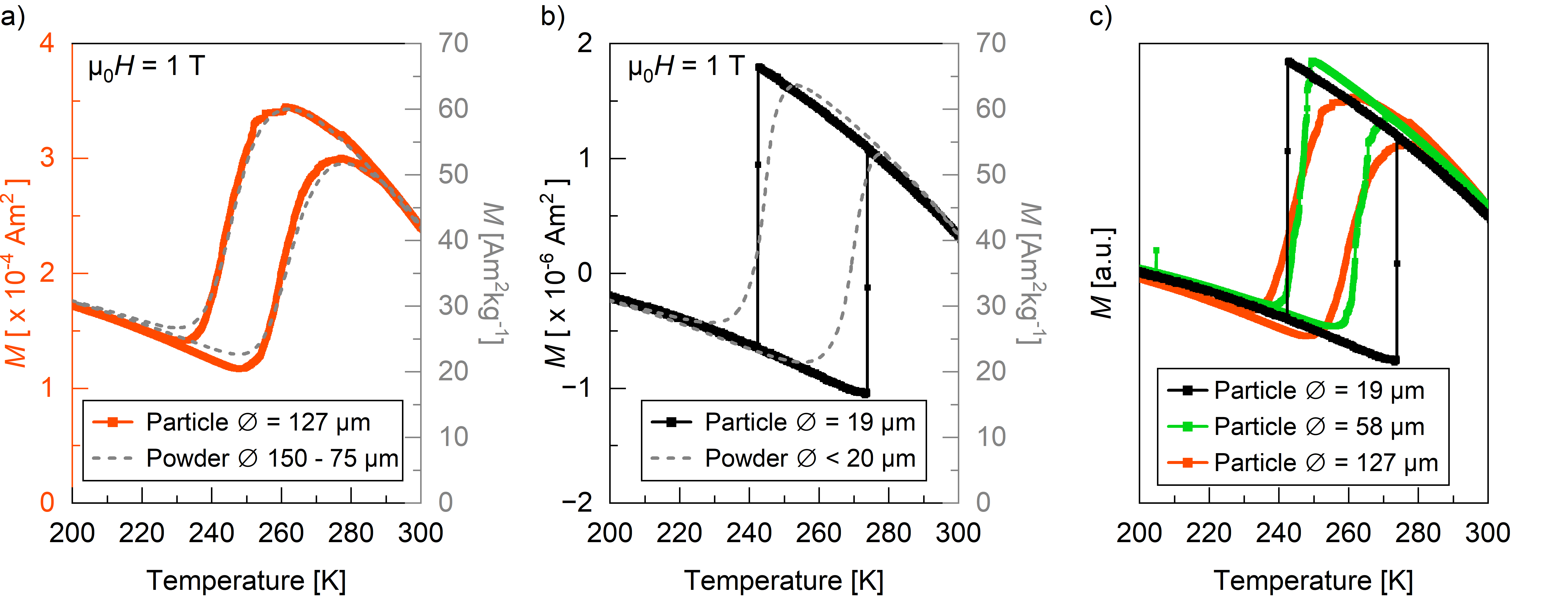}
    \caption{Comparison between $M(T)$-curves of powder and single particle from size fraction a) 150-75\,µm and b) $<$20\,µm. Normalized $M(T)$-curves of individual particles c) with large, small, and intermediate size reveal size-dependent martensitic transformation.}
    \label{fig:comparison}
\end{figure}

\begin{figure}[pos=H]
	\centering
		\includegraphics[width=\textwidth]{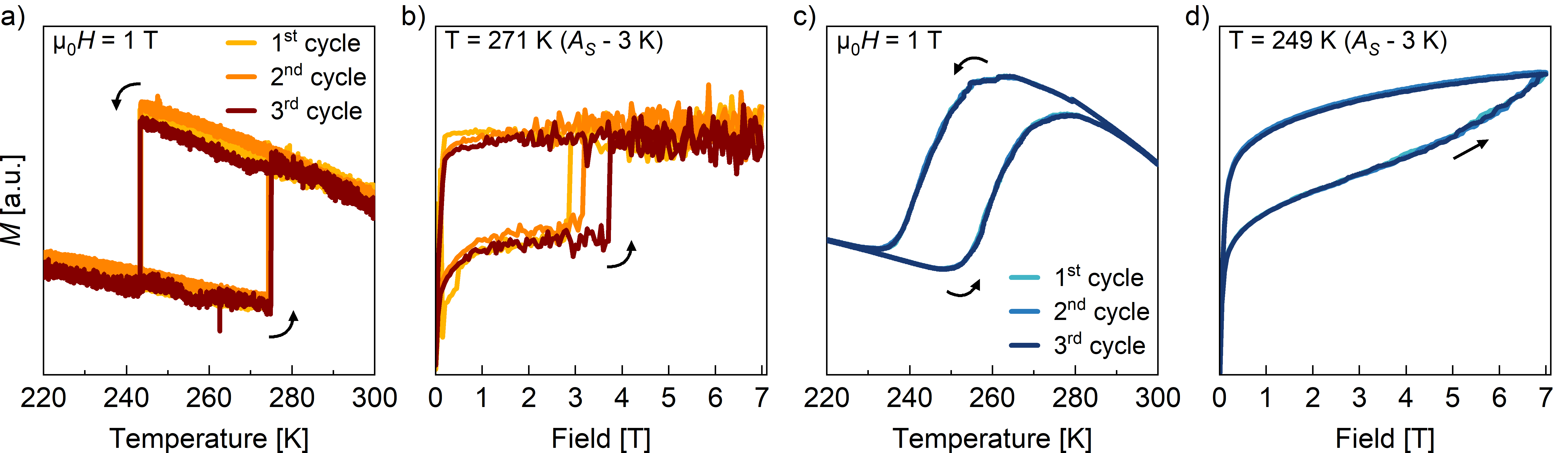}
    \caption{$M(T)$-curves showing the temperature-induced FOMST of a) 19\,µm and c) 127\,µm single particle and $M(H)$-curves showing the field-induced FOMST of b) 19\,µm and d) 127\,µm single particle, recorded for three consecutive cycles, following a heating and cooling protocol between cycles. $M(H)$-curves of b) 19\,µm particle reveal statistical character of field-induced phase transition in small particle.}
    \label{fig:fieldvstemperature}
\end{figure}

Figure \ref{fig:MT} a) shows temperature-dependent magnetization ($M(T)$) curves of all powders in an applied magnetic field of 1\,T between 150\,K and 350\,K under field-cooling and field-heating protocol. The $M(T)$-curves for all powders show a complete, fully reversible FOMST. The $M(T)$-curves and absolute values of the magnetization above and below the FOMST coincide for all size fractions, with $M$ = 37\,Am\textsuperscript{2}kg\textsuperscript{-1} at 150\,K and $M$ = 6\,Am\textsuperscript{2}kg\textsuperscript{-1} at 350\,K. This is consistent with the XRD data, indicating the absence of residual austenite below $M_f$ and residual martensite above $A_f$. Between 200\,K and 290\,K, the FOMST is induced. As for the phase fraction analysis of the XRD data, the magnetization curves also reveal a size-dependent range of transition temperatures and thermal hysteresis. ($M_s - M_f$) and ($A_f - A_s$) decrease steadily with decreasing particle size; the values range between 19\,K for 150-75\,µm and 9\,K for $<$20\,µm, and 20\,K for 150-75\,µm and 11\,K for $<$20\,µm, respectively.  
The inverse trend is observed for $\Delta T_{hyst}$, which gradually increases with decreasing particle size, from 19\,K for 150-75\,µm to 28\,K for $<$20\,µm. 
In contrast to the transition width and thermal hysteresis, no particle size dependence is observed for the transition temperature $T_t$ and the austenite Curie temperature $T_c^A$. As shown in Figure~\ref {fig:MT} b), the variation between different size fractions lies within the error bar of the measurement. As $T_c^A$ and $T_t$ are highly sensitive to the chemical composition \cite{Cakr.2015,Taubel.2018}, constant values indicate no variation in the chemical composition between the different powders and is in agreement with the results from ICP-OES and XRD measurements.
To elucidate the size dependence of the transition widths and thermal hysteresis in more detail, a total of 28 particles between 13\,µm and 127\,µm are selected from the powders and characterized by $M(T)$ in an external field of 1\,T. By performing measurements on individual particles, we can differentiate whether the FOMST characteristics of $\Delta T_{hyst}$, ($ M_s - M_f$) and ($A_f - A_s$) result from the collective measurement of multiple particles or are characteristics of the particle itself.
Figure \ref{fig:comparison} a) and b) exemplary show the $M(T)$-curves of a single particle for a large ($\varnothing$\,=\,127\,µm) and a small ($\varnothing$\,=\,19\,µm) particle, alongside the $M(T)$-curve of the corresponding powder fraction for comparison. Figure \ref{fig:comparison} c) shows a comparison of the two single particles and a particle with an intermediate size ($\varnothing$\,=\,58\,µm). 
Figure \ref{fig:comparison} a) shows almost identical characteristics of the phase transition for the large particle ($\varnothing$\,=\,127\,µm) and the powder. In both cases, a smooth and continuous transformation is observed, with an identical hysteresis of 17\,K and nearly identical transition temperature ranges of ($M_s - M_f$)\,=\,20.3\,K and ($A_f - A_s$)\,=\,17.8\,K for the single particle, and ($M_s - M_f$)\,=\,19.3\,K and ($A_f - A_s$)\,=\,19.6\,K for the powder.
In contrast, the characteristics of the FOMST in the case of the small particle ($\varnothing$\,=\,19\,µm) differ significantly from those of the powder, as shown in Figure \ref{fig:comparison} b).
In the small single particle, a very sharp and abrupt phase transformation is observed, which is in contrast to the continuous transformation behavior found in the corresponding powder and the larger particle.
The transition temperature ranges of ($ M_s - M_f$)\,=\,0.1\,K and ($A_f - A_s$)\,=\,0.1\,K indicate a jump-like phase transformation, in which the whole particle is fully transformed once the nucleation of the FOMST occurs. In contrast, the phase transition in the powder extends over temperature ranges of ($M_s - M_f$)\,=\,8.9\,K and ($A_f - A_s$)\,=\,10.6\,K, which can be explained by a superposition of individual particles with jump-like phase transformation with slight variation in the transition temperature. In contrast, the thermal hysteresis of the single particle is 27\,K, again comparable to the 26\,K determined for the powder. Unlike the transition width, the thermal hysteresis is not affected by the superposition of particles with different transition temperatures, provided that cross-coupling between individual particles is negligible \cite{Waske2015}. 
Figure \ref{fig:comparison} c) shows the normalized $M(T)$-curves of the two particles and an additional particle with an intermediate size ($\varnothing$\,=\,58\,µm). The transformation behavior of the intermediate-size particle resembles a combination of the two previously discussed transformation characteristics. The transformation extends over temperature ranges of ($ M_s - M_f$)\,=\,7.4\,K and ($A_f - A_s$)\,=\,9.6\,K, but occurs in steps, as multiple smaller, jump-like features can be recognized. The hysteresis of 19\,K lies between the small and large particles. This confirms that the FOMST indeed depends on the particle size. $\Delta T_{hyst}$ decreases with increasing particle size, while ($M_s - M_f$) and ($A_f - A_s$) decrease with decreasing particle size. The overlap of the particle and the grain size distribution decreases with increasing particle size and, therefore, the probability of finding larger single-crystalline particles. For particles with a size comparable to or smaller than the average grain size, a jump-like transition becomes more likely. To prove that the particles in Figure \ref{fig:comparison} a) and b) are indeed of single- or oligocrystalline and polycrystalline nature, XRD analysis of individual particles is shown in Figure S6 in the Supplementary Information.\\
To elucidate the dependence of the FOMST on the external stimulus, the temperature-induced transformation is compared with the magnetic field-induced FOMST for both small ($\varnothing$ = 19\,µm) and large ($\varnothing$ = 127\,µm) particles, introduced already in Figure \ref{fig:comparison} a) and b). Each measurement was repeated three times to assess the reproducibility of the phase transition.
For the field-induced transformation, a defined cooling and heating protocol was applied. The samples were cooled from a fully austenitic state at room temperature to a fully martensitic state, followed by heating to the target temperature $A_s\,-$\,3\,K. This ensures comparability, as the initial state is fixed. Figure \ref{fig:fieldvstemperature} shows the a) temperature- and b) field-induced transformation for the small particle, as well as the c) temperature- and d) field-induced transformation for the large particle.
In each case, the temperature-induced FOMST is fully reproducible and occurs at the same temperature for both the forward and reverse transformations, consistently following the same transformation path. In contrast, differences emerge for the field-induced FOMST from martensite to austenite in the small particle. Upon reaching the critical field, the magnetization increases rapidly, and the particle undergoes a jump-like transformation, similar to that observed in temperature-induced FOMST. However, repeated cycling through the phase transition yields different critical fields for each cycle, namely $\mu_0 H_{c1}$\,=\,2.90\,T, $\mu_0 H_{c2}$\,=\,3.20\,T and $\mu_0 H_{c3}$\,=\,3.75\,T. This behavior is confirmed by additional measurements on the same particle as well as on other particles (Figure S3 in the Supplementary Information). No systematic change in the critical field with increasing cycle number is observed. Instead, the critical field fluctuates between consecutive cycles. This behavior is inconsistent with a training effect, which would be expected to result in a progressive shift of the critical field in the same direction with increasing cycle number. This is supported by Figure S3 in the Supplementary Information, where the critical field does not evolve monotonically but rather exhibits irregular cycle-to-cycle variations. Moreover, all particles were pre-cycled and had already undergone the phase transition multiple times prior to the measurements.
These observations indicate that nucleation of FOMST in small particles is stochastic, and that jump-like phase transitions reflect a single transformation event. In contrast, the large polycrystalline particle shows no significant difference between temperature- and field-induced transformations, and the nucleation and growth processes seem to follow the same path, indicating that the microstructure of the particle defines the transformation path and the transformation does not occur randomly.

\subsubsection{Calorimetry}
\begin{figure}[pos=H]
	\centering
		\includegraphics[width=\textwidth]{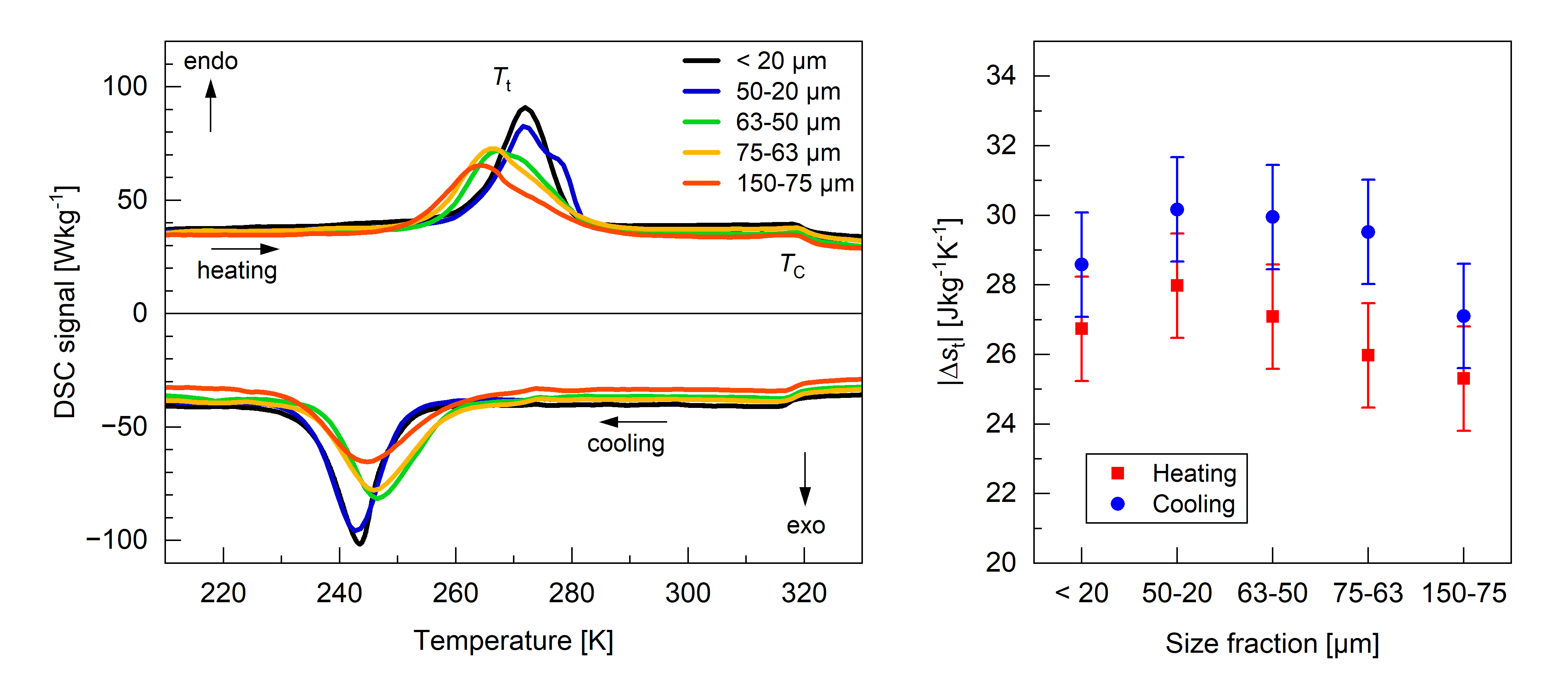}
    \caption{a) DSC curves of all powder size fractions and b) respective transition entropy changes \(\Delta s_t\).}
    \label{fig:DSC}
\end{figure}

Figure \ref{fig:DSC} a) shows the DSC curves of all powders in the temperature range of 210\,K to 330\,K for the heating and cooling protocol. The strong endothermic peaks on heating correspond to the reverse martensitic transformation, while the strong exothermic peaks on cooling correspond to the martensitic transformation. The shoulder in the heating and cooling curve at 319\,K indicates the second-order phase transition from ferromagnetic to paramagnetic austenite and corresponds to $T_c^A$. 
In agreement with previous XRD- and $M(T)$-analysis, the DSC curves also show a dependence on particle size. 
The powder size fraction < 20\,µm shows the highest peak intensity and the narrowest peak width, with an FWHM of 9.6\,K upon heating and 7.7\,K upon cooling. With increasing size, the peak width increases gradually while the peak intensity decreases gradually. The size fraction 150-75\,µm exhibits the lowest peak intensities and the largest FWHM of 16.4\,K upon heating and 15.9\,K upon cooling.
From the DSC signal, the transition entropy change \(\Delta s_t\) is calculated according to Equation (\ref{eq:dS-DSC}) and displayed for the different powder size fractions in Figure \ref{fig:DSC} b). 
Upon heating, \(\Delta s_t\) takes values between 25.3\,J(kgK)\textsuperscript{-1} for size fraction 150-75\,µm and 28.0\,J(kgK)\textsuperscript{-1} for size fraction 50-20\,µm. The results also show that the difference between \(\Delta s_t\) determined from the exothermic and endothermic peaks is small ($<$\,3.6\,J(kgK)\textsuperscript{-1}) and, except for powder 75-63\,µm, in the error bar of each other. In each case, a dependency of \(\Delta s_t\) on the different particle sizes cannot be proven. 

\section{Discussion}
\begin{figure}[pos=H]
	\centering
		\includegraphics[width=\textwidth]{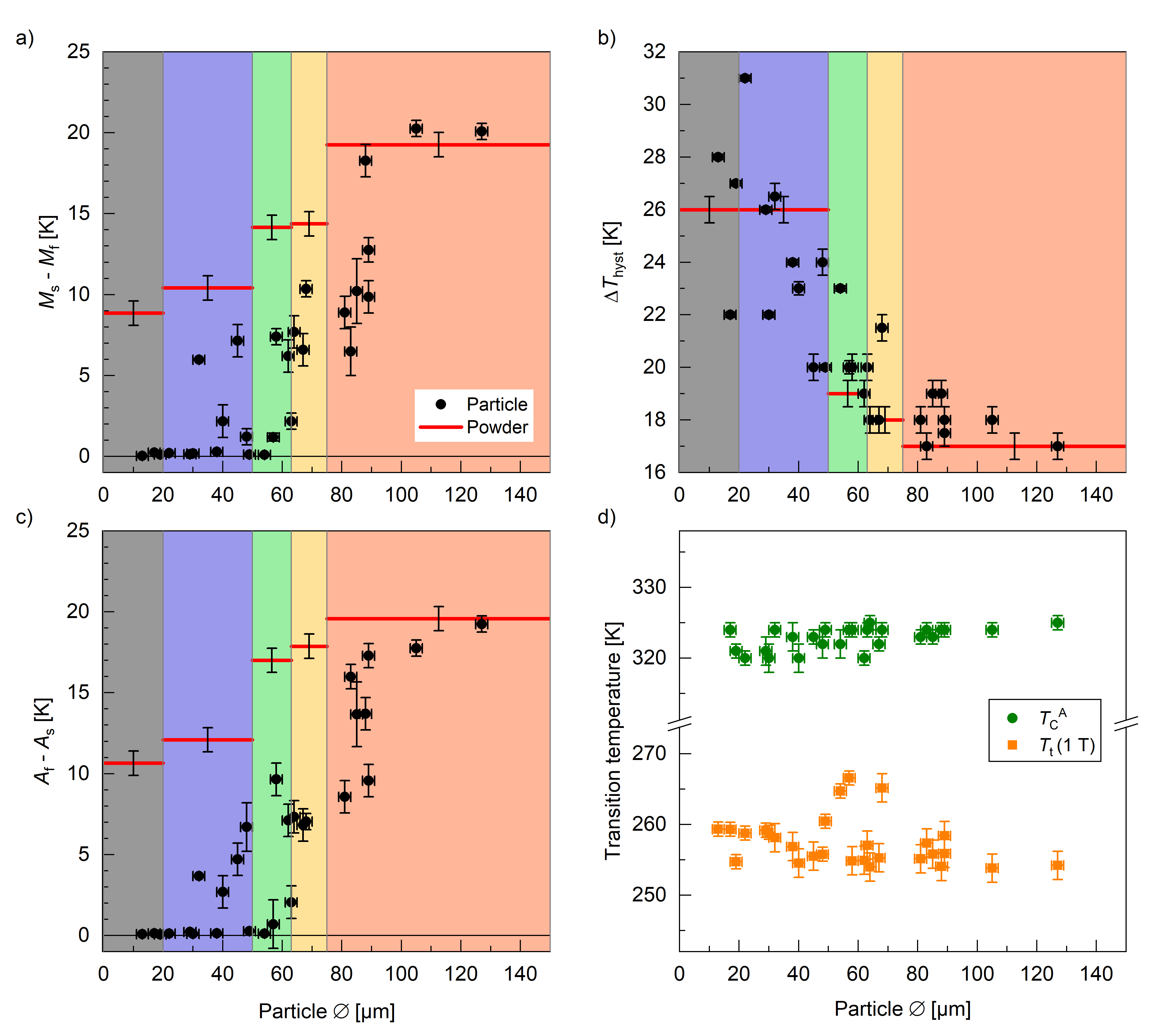}
    \caption{Transition temperature ranges a) ($M_s - M_f$) and c) ($A_f - A_s$), b) thermal hysteresis $\Delta T_{hyst}$, as well as d) Curie temperature $T_c^A$ and transition temperature $T_t$ of all powder fractions and single particles as a function of particle size.}
    \label{fig:Sizeeffect}
\end{figure}

\begin{figure}[pos=H]
	\centering
		\includegraphics[width=\textwidth]{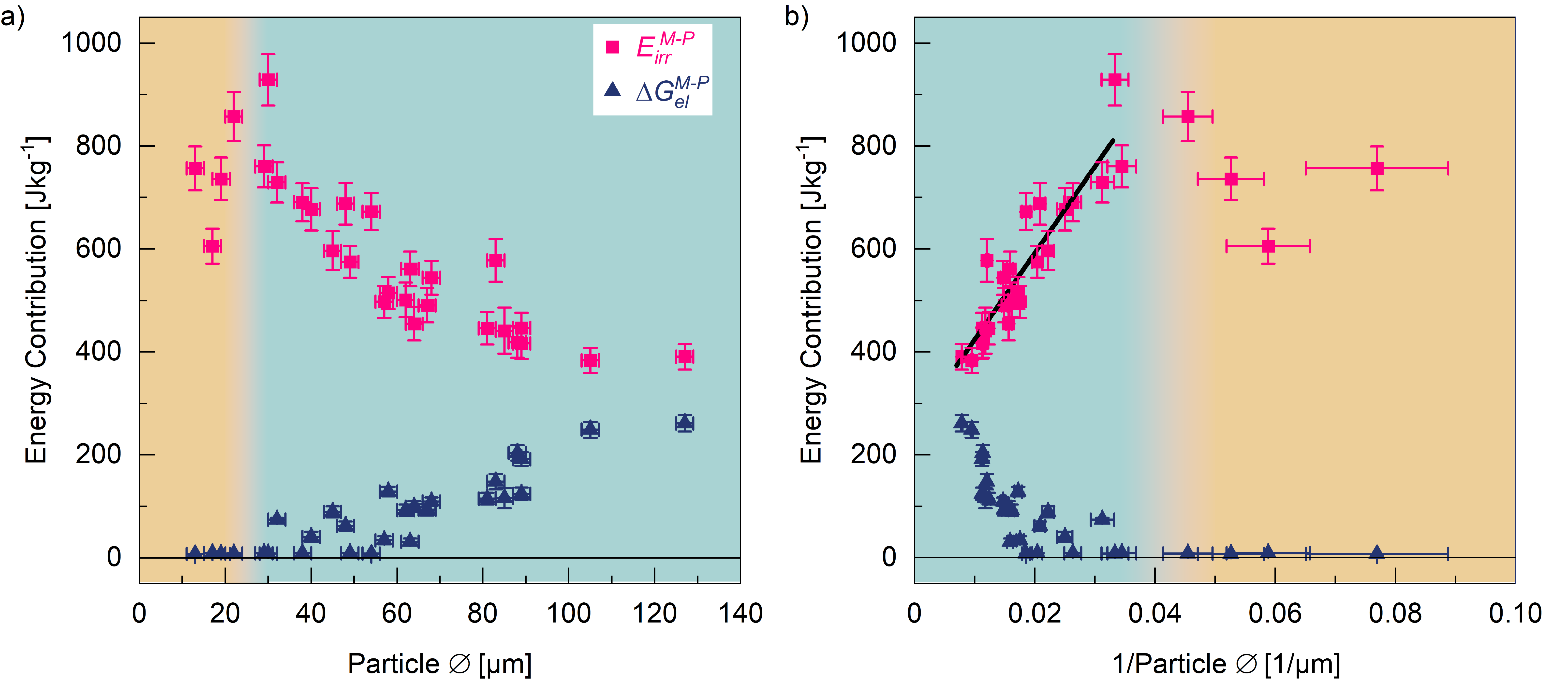}
    \caption{a) Calculated reversible elastic energy $\Delta G_{el}^{M-P}$ and irreversible energy losses $E_{irr}^{M-P}$ for every single particle according to the model of thermoelastic martensitic phase transformations; b) Plot over inverse particle diameter reveals a linear dependence of $E_{irr}^{M-P}$.}
    \label{fig:Energy}
\end{figure}

We could show that a pronounced particle size dependence of the FOMST is consistently observed across XRD, DSC, and magnetometry measurements. EBSD analysis (Figure \ref{fig:EBSD_GrainSize}) indicates that the grain size remains constant across all particle size fractions, whereas the number of grains per particle increases with particle size. Consequently, the smallest particles are predominantly of single-crystalline or oligocrystalline nature, while larger particles are polycrystalline. Temperature-dependent magnetometry of powder fractions captures the overall size-dependent trends, whereas measurements on individual particles elucidate the underlying transformation behavior.
Figure \ref{fig:Sizeeffect} summarizes the particle size dependence of FOMST properties for both individual particles and loose powder fractions, including a) ($M_s - M_f$), b) $\Delta T_{hyst}$, c) ($A_f - A_s$), and d) $T_t$ and $T_c^A$. 
($M_s - M_f$) and ($A_f - A_s$) decrease with decreasing particle size for both powder and individual single particles. For particle sizes larger than 100\,µm, the transition ranges of single particles lie within the range observed for the powder. Below 100\,µm, ($M_s - M_f$) and ($A_f - A_s$) decrease more rapidly for the single particles, dropping almost to zero when the particle size reaches the single-grain limit for $\varnothing \approx 20-30$\,µm. For these particles, an abrupt, jump-like phase transformation of the entire particle is observed. Such abrupt transformations have also been reported for individual Ni-Co-Mn-Sn and Ni-Co-Mn-In particles by Zhang \textit{et al.} \cite{Zhang.2021} and Lago \textit{et al.} \cite{Lago.2024}, respectively. Zhang \textit{et al.} observed both abrupt and gradual transformations among particles of similar size, consistent with the variability in the intermediate size range of the present single-particle measurements. Particle size alone therefore does not determine the transformation pathway, differences in microstructure may also contribute. For the powder, ($M_s - M_f$) and ($A_f - A_s$) show a remaining width of approximately 10\,K, which originates from the superposition of multiple $M(T)$ curves from single particles with jump-like phase transformation but shifted transition temperatures $M_s$, $M_f$, $A_f$, and $A_s$. The variation in the transition temperature $T_t$, shown in d), can result from small variations in chemical composition among the particles, leading to an artificial broadening of the FOMST determined in powder samples. The variation in $T_t$ and $T_c^A$ is present across the entire particle size range and shows no trend with particle size. $T_c^A$ ranges between 320\,K and 325\,K, while $T_t$ varies between 254\,K and 267\,K. As the size of the single particles increases, ($M_s - M_f$) and ($A_f - A_s$) increase accordingly and the artificial broadening of the FOMST determined in powder due to chemical variation among the particles becomes less dominant and can be neglected for particles above 100\,µm. 
The thermal hysteresis $\Delta T_{hyst}$ decreases with increasing particle size, approaching a minimum of approximately 17\,K for particles above 100\,µm. For particle sizes above 50\,µm, $\Delta T_{hyst}$ in individual particles is comparable to that in powder, whereas for smaller particles $\Delta T_{hyst}$ shows a large fluctuation among the individual particles around the corresponding powder values.
Unlike ($M_s - M_f$) and ($A_f - A_s$), a variation in $T_t$ does not lead to an artificial broadening of $\Delta T_{hyst}$. Zhang \textit{et al.} \cite{Zhang.2021} likewise reported increasing thermal hysteresis with decreasing particle size. They associated this trend with the lower probability of highly potent nucleation sites in smaller volumes and, for gradually transforming particles, with greater frictional dissipation during interface motion. Both effects could contribute to the trend observed here, although their relative contributions cannot be determined from the present measurements.\\
In large particles, the combination of polycrystallinity and preferential nucleation at the particle surface seems to facilitate the initiation of the martensitic transformation by lowering undercooling and overheating, shifting $M_s$ to higher and $A_s$ to lower temperatures and thus, reduce thermal hysteresis. At the same time, grain boundaries impede the propagation of the phase boundary, resulting in a gradual transformation, broader transition ranges, and shifts of $M_f$ and $A_f$ to lower and higher temperatures, respectively. A related observation was reported by Taubel \textit{et al.} \cite{Taubel.2020} for bulk Ni-Co-Mn-Ti, where a chemically homogenized sample with exceptionally large grains exhibited a sharp FOMST and the authors proposed that the lower density of grain boundaries reduces barriers to martensite growth. In contrast, the reduced number or absence of grain boundaries in smaller particles is consistent with a higher nucleation barrier, as the entire particle transforms in a single event, leading to an increase in $\Delta T_{hyst}$ and a jump-like transformation with vanishing transition width. This interpretation is supported by temperature-dependent in-situ microscopy (Figure S4 and S5 in the Supplementary Information) on particles from the $<$20\,µm and 150-75\,µm size fractions. Small, single-crystalline particles exhibit rapid, jump-like transformations immediately upon nucleation, whereas larger polycrystalline particles transform continuously over a broader temperature interval. In each case, the FOMST is initiated by preferential nucleation of martensite at the particle surface. The larger variation in thermal hysteresis observed among small particles may further indicate the importance of individual defects and preferential nucleation sites. Large particles contain a comparatively large number of potential nucleation sites, such that differences between individual particles are averaged out and their behaviour approaches that of the powder ensemble. In contrast, the number of preferential nucleation sites may vary substantially between small particles, resulting in stronger particle-to-particle variations in the thermal hysteresis. 
Field-dependent magnetometry (Figure S2 in Supplementary Information) shows no size dependence of the magnetization process or the absolute magnetization. The change in transition entropy, as determined by DSC, is also independent of particle size.\\
To evaluate the effect of particle size within the framework of thermoelastic martensitic transformation, we used the model developed by Ortín and Planes \cite{Ortin.1989} to correlate the particle size to the contributions of reversible elastic energy and irreversible energy losses. In this model, the reverse phase transition from martensite to the parent austenite phase (M–P) proceeds under conditions of local Gibbs free energy equilibrium $\Delta G^{M - P}$. This equilibrium condition can be expressed as a balance among the chemical driving force, the elastic energy contribution, and the irreversible energy losses:
    \begin{equation}
        \Delta G^{M - P} = \Delta G^{M - P}_{ch}\:-\: \Delta G^{M - P}_{el} \:+\: E^{M - P}_{irr}\:=\:0 
        \label{eq:dG}
    \end{equation}
Here, the term $\Delta G^{M - P}_{ch}$ corresponds to the difference in Gibbs free energy between the martensite and austenite. This term acts as the primary driving force for the transformation and depends explicitly on temperature and applied magnetic field. The elastic contribution $\Delta G^{M - P}_{el}$ represents the reversible elastic strain energy stored during the transformation. This contribution arises from interfacial effects at various boundaries, for example, habit planes between martensite and austenite and grain boundaries, in addition to elastic strains caused by lattice distortions.
According to Shamberger and Ohuchi \cite{Shamberger.2009}, the elastic energy $\Delta G^{M - P}_{el}$ and irreversible losses $E^{M - P}_{irr}$ can be approximated as:
\begin{equation}
    \Delta G^{M - P}_{el} = \frac{1}{2} \Delta s_t \Delta T_{el} \quad with \quad \Delta T_{el} = [(A_f - A_s) + (M_s - M_f)]/2  
    \label{eq:G_el}
\end{equation}
\begin{equation}
     E^{M - P}_{irr} = \Delta s_t \Delta T_{irr} \quad with \quad \Delta T_{irr} = [(A_f + A_s) - (M_s + M_f)]/2 
     \label{eq:E_irr}
\end{equation}
where $\Delta T_{el}$ is related to the width of the phase transition, and $\Delta T_{irr}$ represents the average thermal hysteresis of the material.\\
To evaluate the effect of particle size on the energy contributions $\Delta G_{el}^{M-P}$ and $E_{irr}^{M-P}$ we calculated them for every single particle using Equations (\ref{eq:G_el}) and (\ref{eq:E_irr}) using $M_s$, $M_f$, $A_s$ and $A_f$ measured at 1\,T and extrapolated to zero field, together with the entropy change obtained from DSC. Figure \ref{fig:Energy} presents $\Delta G_{el}^{M-P}$ and $E_{irr}^{M-P}$ as a function of a) particle diameter and b) inverse particle diameter. 

The irreversible energy losses increase strongly with decreasing particle size, reaching values up to approximately twice those of the largest particles. According to the thermodynamic model, $E_{irr}^{M-P}$ scales with the thermal hysteresis, which likewise increases with decreasing particle size. This behaviour can be related to particle-size-dependent nucleation. Grain boundaries, particularly triple junctions, may act as preferential nucleation sites because their associated local stress fields can reduce the nucleation energy barrier, shifting the transition start temperatures ($M_s$ and $A_s$) closer to the equilibrium transformation temperature. Large polycrystalline particles contain numerous such potential nucleation sites and therefore exhibit comparatively low irreversible losses and thermal hysteresis. In contrast, small, predominantly single-grain particles contain fewer preferential nucleation sites and may consequently require larger undercooling or overheating for nucleation, resulting in increased thermal hysteresis and irreversible energy losses.
The elastic energy contribution shows the opposite particle-size dependence. It is highest for large particles and approaches zero for several of the smallest particles. Within the thermodynamic model, $\Delta G_{el}^{M-P}$ scales with the transformation width, which decreases with decreasing particle size and becomes very small for the smallest particles. Elastic strain energy accumulates during the FOMST as a consequence of the accommodation of transformation strains, including interactions between martensitic variants and constraints imposed by grain boundaries and other internal interfaces. In large polycrystalline particles, these interfaces can hinder the progression of the habit plane and promote a gradual transformation over a broader temperature interval, resulting in larger transformation widths and higher elastic energy contributions. In small, single-grain particles, the reduced number of internal interfaces limits these constraints and interactions between differently oriented martensitic variants. In addition, transformation-induced strains can be accommodated more readily at the free particle surface. Both effects are consistent with reduced accumulation of elastic strain energy. Accordingly, once nucleation occurs, these particles can undergo an almost instantaneous, jump-like transformation in which the entire particle transforms in a single event.
Figure \ref{fig:Energy} b) further shows $\Delta G_{el}^{M-P}$ and $E_{irr}^{M-P}$ as a function of inverse particle diameter. The irreversible energy losses exhibit an approximately linear dependence on inverse particle diameter, as indicated by the linear fit. For gradually transforming Ni-Co-Mn-Sn particles classified as single crystals, Zhang \textit{et al.} \cite{Zhang.2021} reported a comparable size dependence of thermal hysteresis, $\Delta T_{hyst}\propto(r^*)^{-0.9 \pm 0.1}$, where $r^*=r/3$ for a spherical particle of radius $r$. The approximately inverse-diameter dependence in the present study is observed among a larger particle size range and individual particles that also include abruptly transforming small particles and larger polycrystalline particles. Because the transition entropy remains nearly constant across the present powder size fractions, the size dependence of $E_{irr}^{M-P}$ reflects that of thermal hysteresis, which is attributed to the pinning effect of defects at the particle surface. The smallest particles deviate from the approximately linear trend, although the overall increase in irreversible energy losses with decreasing particle size remains evident. This deviation may partly reflect uncertainty in determining the size of irregularly shaped particles. Differences in the availability of potent nucleation sites may also increase particle-to-particle variability at smaller volumes. Lago \textit{et al.} \cite{Lago.2024} reported a wide range of undercooling among Ni-Co-Mn-In particles with volumes below approximately 5000\,µ$\mathrm{m}^3$ (equivalent to a diameter of approximately 21 µm for a sphere), which they attributed to differences in the availability of potent nucleation sites among similarly sized particles. Such variability could contribute to the scatter observed here, although their undercooling and the present irreversible energy contribution are distinct measures. Furthermore, Ueland et al. reported that the stress hysteresis of the stress-induced martensitic transformation in Cu-based shape-memory alloy microwires scales with $1/D$, where $D$ is the wire diameter \cite{Ueland.2013}. They attributed this dependence to surface-controlled behavior. Below approximately 10 µm however, they described a transition to obstacle-starvation-controlled behavior, accompanied by a decrease in hysteresis. They attributed this decrease to fewer defects encountered by the moving austenite–martensite interface per unit volume, resulting in reduced frictional energy dissipation. A similar reduction in frictional energy dissipation could tentatively explain the trend in $E_{irr}^{M-P}$, which scales directly with $\Delta T_{hyst}$, observed in the present study. In contrast, the elastic energy contribution decreases with inverse particle diameter and does not exhibit a clear linear dependence.

\section{Conclusion}
We systematically investigated the influence of particle size and microstructure on the FOMST in gas-atomized Ni-Mn-Sn Heusler alloy powder. By combining microstructural characterization with temperature-dependent XRD, magnetometry, calorimetry, and single-particle measurements, the individual contributions of particle size and crystallinity to the transformation characteristics were disentangled.

A pronounced particle size dependence of the FOMST was identified. With decreasing particle size, the transition ranges ($M_s - M_f$) and ($A_f - A_s$) continuously narrow, whereas $\Delta T_{hyst}$ increases. These trends are consistently observed across both powder fractions and individual particles. The transition entropy change, the mean transition temperature $T_t$, and intrinsic magnetic properties, including saturation magnetization and Curie temperature of the austenite, remain unaffected. 

EBSD analysis reveals a transition from polycrystalline particles at larger sizes to predominantly single-crystalline particles below $\varnothing \approx 20-30$\,µm. This change in microstructure is identified as the key factor governing the observed size dependence. In polycrystalline particles, the presence of grain boundaries, combined with free interfaces of the surface grains, serves as a preferential nucleation site, enabling a gradual transformation over extended temperature ranges and reducing thermal hysteresis. In contrast, the reduced number or absence of grain boundaries in smaller particles hinders the nucleation and the particle is forced to transform in one nucleation event, resulting in abrupt, jump-like transformations with minimum transition ranges and increased hysteresis, as observed in single-particle magnetometry measurements. 

We evaluated the energy contributions to the FOMST within the framework of thermoelastic martensitic transformations. The reversible elastic energy $\Delta G_{el}^{M-P}$ decreases with decreasing particle size due to narrowing transition ranges, whereas the irreversible energy losses $E_{irr}^{M-P}$ increase and scale approximately with the inverse particle diameter, reflecting larger $\Delta T_{hyst}$ and higher nucleation barriers in smaller particles.

Overall, these results demonstrate that the size dependence of the FOMST in Ni-Mn-Sn is primarily governed by the transition from polycrystalline to single-crystalline particles and the associated change in nucleation and growth behavior. This provides a framework for understanding size effects in magnetocaloric Heusler alloys and highlights the importance of controlling particle size and microstructure for optimizing transformation characteristics in powder-based applications. Specifically, these results identify an application-dependent trade-off between the sharpness of the FOMST and thermal hysteresis. Small single-crystalline particles may be suitable for sensing applications requiring an abrupt magnetic response, although their larger hysteresis and switching reproducibility must be considered. Larger polycrystalline particles exhibit lower thermal hysteresis and broader transformation ranges. Their lower hysteresis may be advantageous for cyclic magnetocaloric operation in minor loops when the available magnetic-field change is limited.

\medskip
\textbf{Acknowledgements} \par
The authors acknowledge financial support from the Deutsche Forschungsgemeinschaft (DFG, German Research Foundation) CRC/TRR 270 “HoMMage” (Project-ID 405553726-TRR 270), projects B11, B04, A08 and B01. FS acknowledges funding from the DFG Project-ID 527201505. The authors also thank Konrad Opelt (Fraunhofer IWKS) for performing the particle size analysis and Sebastian Fähler (Helmholtz-Zentrum Dresden-Rossendorf) for fruitful discussion. We acknowledge DESY (Hamburg, Germany), a member of the Helmholtz Association HGF, for the provision of experimental facilities. Parts of this research were carried out at PETRA III beamline P02.1. Beamtime was allocated by an In-House contingent.

\nolinenumbers

\appendix
\section{Supplementary Material}
The supplementary material includes further information on the particle and grain-size distributions, EDX spectra, field-dependent magnetization measurements, single-particle XRD analysis, and in situ microscopy analysis.

\printcredits

\medskip
\bibliographystyle{elsarticle-num}

\bibliography{refs_new}
\label{LastPageMain} 

\clearpage
\end{document}


\makeatletter
\renewcommand \thesection{S\@arabic\c@section}
\renewcommand\thetable{S\@arabic\c@table}
\renewcommand \thefigure{S\@arabic\c@figure}
\makeatother

\title{Supplementary - First-order magnetostructural transition in Ni-Mn-Sn powder - A matter of particle and grain size}

\maketitle
\linenumbers 

\section{Particle size distribution and grain size}

\begin{table}[H]
\centering
\caption{Median D\textsubscript{10}, D\textsubscript{50}, D\textsubscript{90}, average and median grain size of each powder.}
\label{tab:grain_size}

\begin{tabular}{c 
S[table-format=3.1] 
S[table-format=3.1] 
S[table-format=3.1] 
S[table-format=2.1] 
S[table-format=2.1]}
\toprule
Particle size [\si{\micro\meter}] & 
{$D_{\text{10}}$ [\si{\micro\meter}]} & 
{$D_{\text{50}}$ [\si{\micro\meter}]} & 
{$D_{\text{90}}$ [\si{\micro\meter}]} & 
{Average GS [\si{\micro\meter}]} & 
{Median GS [\si{\micro\meter}]} \\
\midrule
150--75  & 79.2 & 108.0 & 148.0 & 19.0 & 9.0 \\
75--63   & 60.8 & 74.3  & 91.2  & 13.4 & 6.7 \\
63--50   & 50.2 & 63.9  & 82.0  & 14.0 & 7.4 \\
50--20   & 17.4 & 33.3  & 55.8  & 14.8 & 10.2 \\
$<$20    & 6.7  & 14.7  & 31.1  & 8.1  & 3.8 \\
\bottomrule
\end{tabular}
\end{table}

\section{Energy dispersive X-ray spectroscopy}
\begin{figure}[H]
	\centering
		\includegraphics[width=\linewidth]{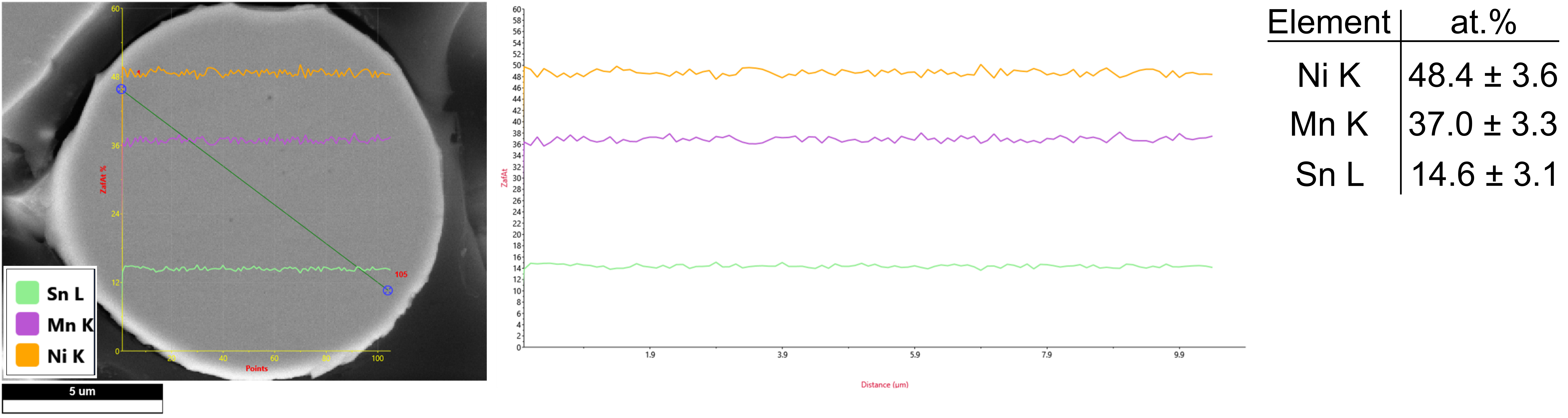}
    \caption{EDX line scan on single particle ($\varnothing$ = 12\,µm) reveals no compositional variation over whole particle cross section.}
    \label{fig:EDX}
\end{figure}

Figure \ref{fig:EDX} shows an EDX line scan across a single particle ($\varnothing$ = 12\,µm). No compositional variation is observed across the particle cross-section.

\section{Field-dependent magnetization curves for different powder size fractions}
\begin{figure}
    \centering
    \includegraphics[width=\linewidth]{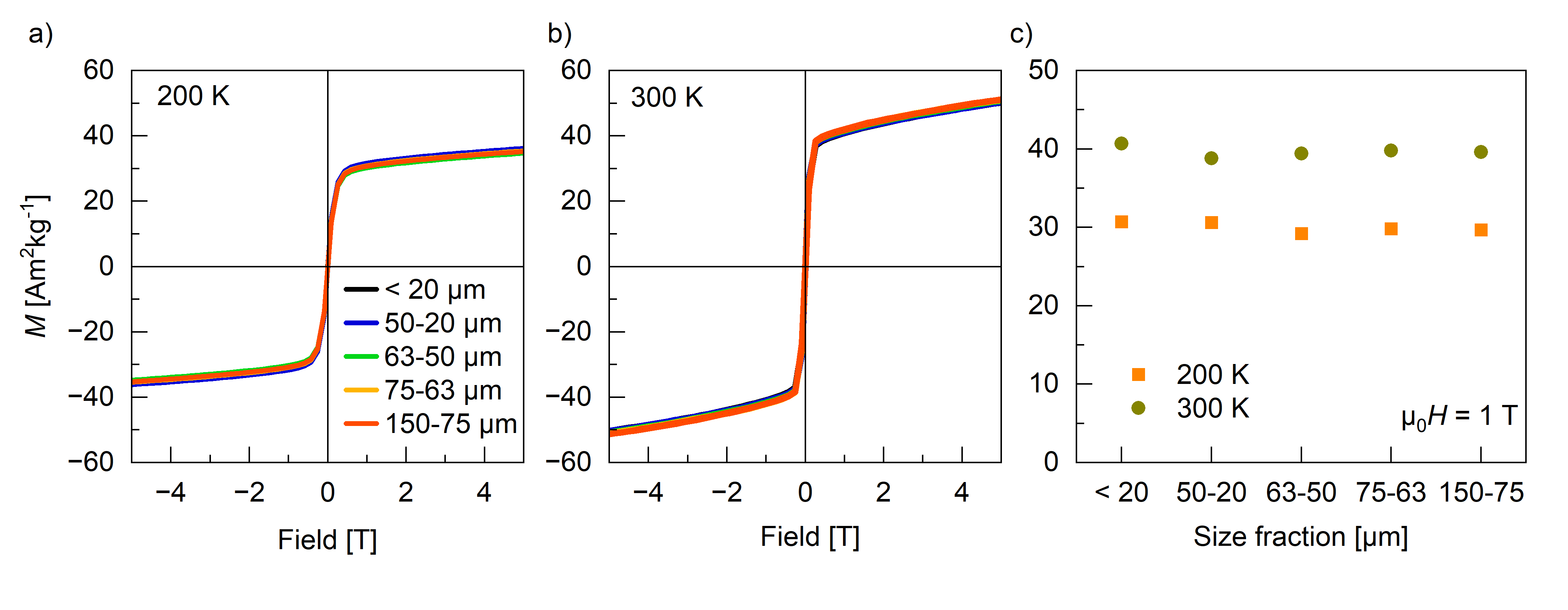}
    \caption{Isothermal magnetization curves of all size fractions under magnetic fields up to 5\,T in the fully martensitic state at a) 200\,K and in the fully austenitic state at b) 300\,K. c) Magnetization value extracted at 1\,T is irrespective of particle size.}
    \label{fig:MH}
\end{figure}

Figure \ref{fig:MH} a) and b) show \textit{M(H)}-curves in the weak magnetic martensite phase at 200 K and in the ferromagnetic austenite phase at 300 K. What can be clearly seen in the \textit{M(H)}-curves is that neither in the martensite phase nor in the austenite phase a difference between size fractions appears. 
When an external field is applied, the magnetization increases rapidly up to roughly 0.5\,T at 200\,K. After that, the magnetization does not reach saturation, but increases linearly, following a slight paramagnetic behavior.
At 300\,K, the magnetization increases more rapidly compared to the behavior at 200\,K, following a linear, paramagnetic behavior from external fields of 0.25\,T on. This non-saturating behaviour could be explained by the close proximity of the measurement temperature and the Curie temperature.
The saturation magnetization at 1\,T is determined for all size fractions and displayed in Figure \ref{fig:MH}c). Values between 29.2 and 30.8\,Am\textsuperscript{2}kg\textsuperscript{-1} at 200\,K in martensite and between 38.8 and 40.7\,Am\textsuperscript{2}kg\textsuperscript{-1} at 300\,K in austenite appear. 

\section{Field-dependent magnetization curves for single particle ($\varnothing$ = 19\,µm)}

\begin{figure}
	\centering
		\includegraphics[scale=.75]{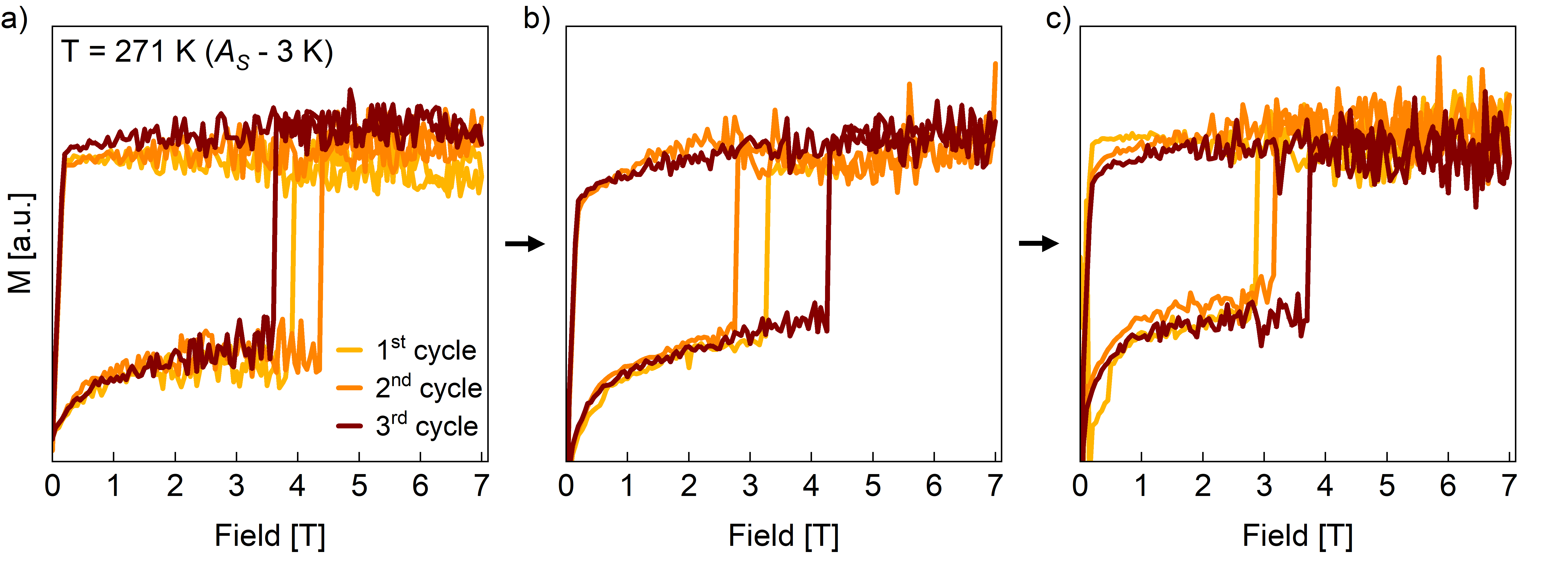}
    \caption{$M(H)$-curves of 19 µm single particle up to 7\,T at 271 K showing non-reproducibility of field-induced FOMST and fluctuation of critical field. a) Initial orientation, b) rotated by 90° and c) repeated for initial orientation. $M(H)$-curves are recorded for three consecutive cycles, following a heating and cooling protocol between cycles.}
    \label{fig:MH_particle_3x}
\end{figure}

As shown in Figure 6 b), the jump-like field-induced FOMST in a small, single-crystalline particle ($\varnothing$ = 19\,µm) is not fully reproducible and shows significant fluctuations in the critical field. Additional $M(H)$-curves for the same particle, presented in Figure \ref{fig:MH_particle_3x}, confirm this behavior.

The initial measurement (Figure \ref{fig:MH_particle_3x} a)) shows critical fields of $\mu_0 H_{c1} = 3.90$ T, $\mu_0 H_{c2} = 4.35$ T, and $\mu_0 H_{c3} = 3.60$ T. To assess a possible influence of anisotropy or particle orientation, the particle was remounted on the measurement stick with a 90° rotation relative to its initial orientation. The corresponding measurements (Figure \ref{fig:MH_particle_3x} b)) yield critical fields of $\mu_0 H_{c1} = 3.25$\,T, $\mu_0 H_{c2} = 2.75$\,T, and $\mu_0 H_{c3} = 4.25$\,T. Subsequently, the particle was remounted to its original orientation and remeasured (Figure \ref{fig:MH_particle_3x} c), resulting in critical fields of $\mu_0 H_{c1} = 2.90$\,T, $\mu_0 H_{c2} = 3.20$\,T, and $\mu_0 H_{c3} = 3.75$\,T. As these measurements were performed after multiple transformation cycles, a training effect can be excluded. Overall, the critical field varies significantly within the range $\mu_0 H_{c,\mathrm{min}} = 2.75$ T to $\mu_0 H_{c,\mathrm{max}} = 4.25$ T.

\section{Characterization of FOMST by in-situ microscopy}

\begin{figure}
	\centering
		\includegraphics[scale=.75]{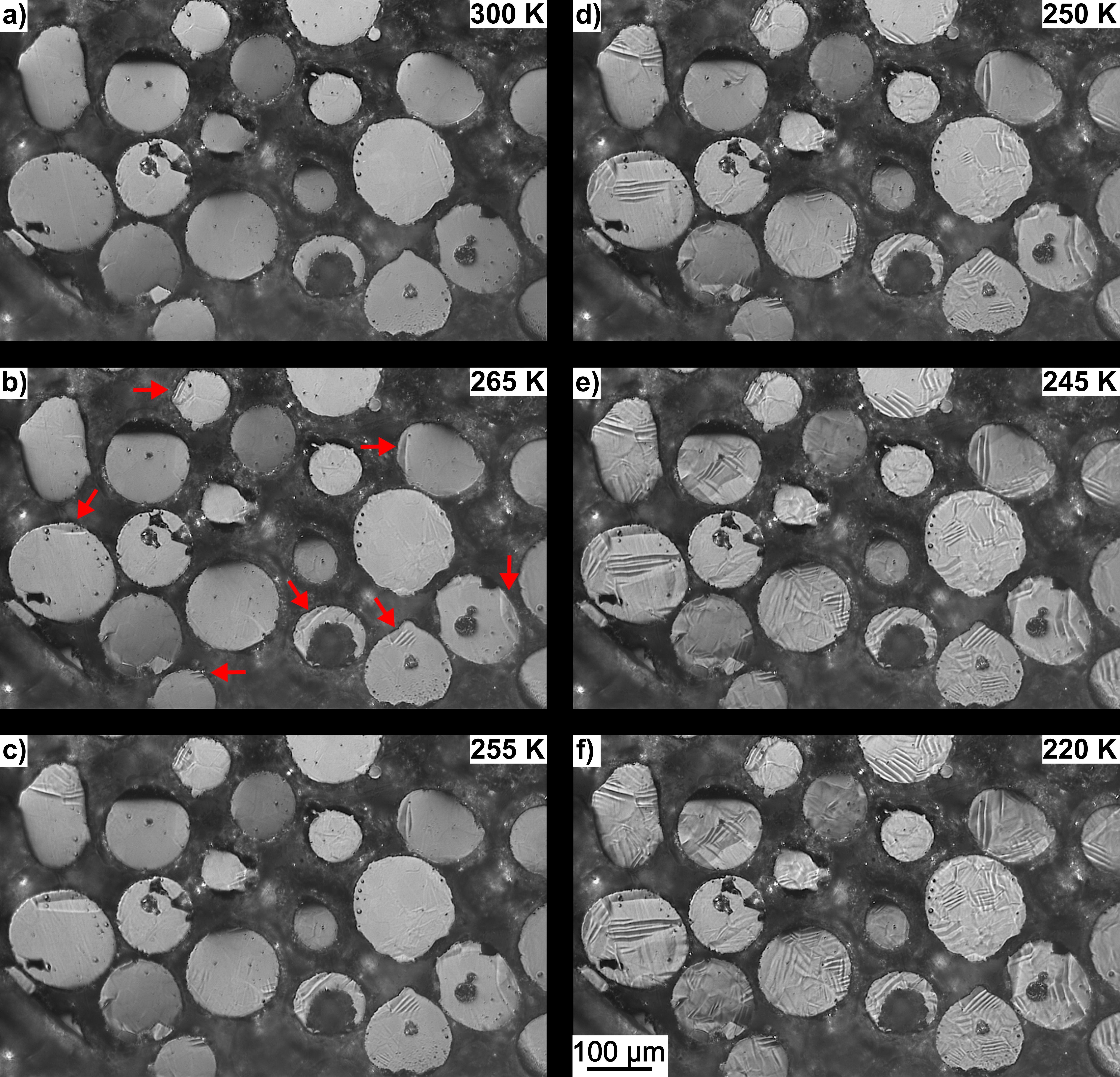}
    \caption{Evolution of martensitic microstructure during first-order magnetostructural transition in Ni-Mn-Sn powder of size fraction 150–75\,µm. Figures are captured by in-situ optical microscopy for temperatures between a) 300\,K and f) 220\,K.}
    \label{fig:In-situ-OM-PC-particle}
\end{figure}

To provide microstructural insight into the martensitic transformation, in-situ optical microscopy was performed on Ni–Mn–Sn particles (150–75 µm) during stepwise cooling from 300\,K to 200\,K. Representative micrographs are shown in Figure \ref{fig:In-situ-OM-PC-particle} at selected temperatures.

At 300\,K (Figure \ref{fig:In-situ-OM-PC-particle} a)), all particles are in the austenitic state. Upon cooling, the martensitic transformation initiates at 265\,K (Figure \ref{fig:In-situ-OM-PC-particle} b)), in agreement with magnetometry data. Martensite formation is observed in multiple particles and is characterized by the appearance of twinned contrast. Nucleation occurs predominantly at particle surfaces and edges. With further cooling (Figure \ref{fig:In-situ-OM-PC-particle} c)–e)), martensite grows into the particle interior, while additional nuclei form at the surface. In some cases, multiple variants nucleate at different surface locations within a single particle, leading to impingement and the formation of macroscopic twin boundaries.

Between 255\,K and 245\,K (Figure \ref{fig:In-situ-OM-PC-particle} c)–e)), the most pronounced microstructural changes are observed. At 255\,K, martensite is mainly confined to surface regions, whereas at 245\,K a significant fraction of the particle volume has transformed, with residual austenite remaining primarily in the centre. A particle-to-particle variation in the onset temperature of martensite formation is evident.

Further cooling to 220\,K (Figure \ref{fig:In-situ-OM-PC-particle} f)) results in an almost fully martensitic microstructure, with only minor residual changes. The martensite exhibits a range of plate sizes and orientations, consistent with a polycrystalline microstructure without preferred texture.

Although the spatial resolution is limited, the observations consistently indicate that martensite nucleation preferentially occurs at free particle surfaces rather than at internal defects. The transformation proceeds via continuous nucleation and growth over a temperature interval of approximately 20–25\,K, in agreement with magnetometry.

\begin{figure}
	\centering
		\includegraphics[scale=.75]{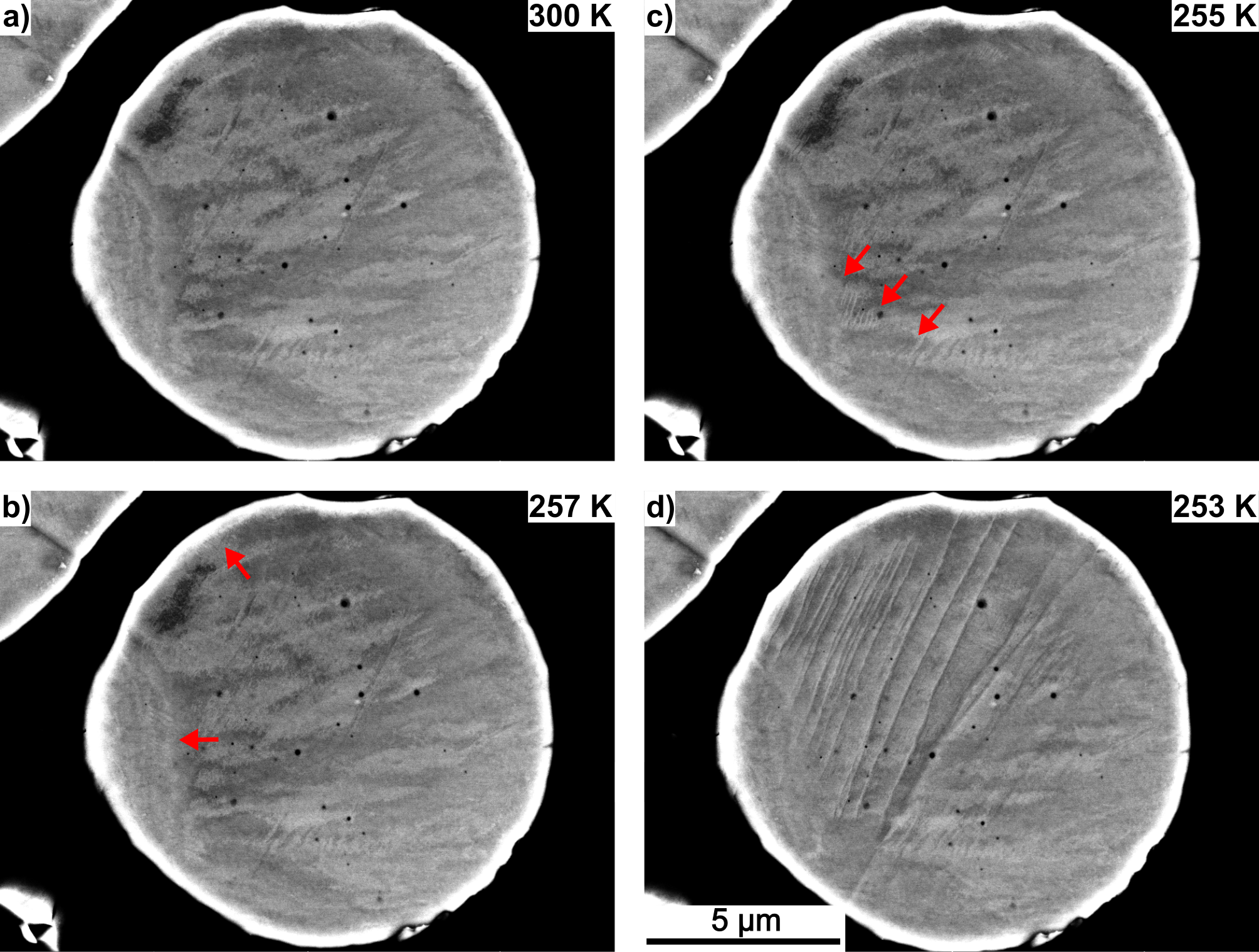}
    \caption{Evolution of martensitic microstructure during first-order magnetostructural transition in Ni-Mn-Sn particle of size fraction $<$ 20 µm. Figures are captured by in-situ BSE-SEM for temperatures between a) 300 K and d) 253 K.}
    \label{fig:In-situ-OM-SC-particle}
\end{figure}

To resolve the transformation behavior in small particles, in-situ SEM was performed on particles with $\varnothing$ $<$20 µm. Representative micrographs of a 12\,µm particle are shown in Figure \ref{fig:In-situ-OM-SC-particle}.

At 300\,K (Figure \ref{fig:In-situ-OM-SC-particle} a)), the particle exhibits no visible grain boundaries and is therefore considered single crystalline. Submicron pores are present throughout the particle, consistent with observations in larger particles and attributed to gas atomization. Variations in contrast are observed but are not associated with compositional differences, as confirmed by EDX, and are therefore attributed to preparation artefacts.

Upon cooling to 257\,K (Figure \ref{fig:In-situ-OM-SC-particle} b)), the first martensitic features appear as nanoscale, needle-like structures, predominantly originating at the particle surface. With further cooling to 255\,K (Figure \ref{fig:In-situ-OM-SC-particle} c)), these nuclei grow into the particle interior, and the advancing phase boundary is evident from the evolving contrast associated with martensite variants.

A rapid, jump-like transformation occurs at 253\,K (Figure \ref{fig:In-situ-OM-SC-particle} d)), during which the particle transforms almost completely and multiple martensite lamellae are formed. The sequence suggests initial nucleation at the surface, followed by limited growth, which subsequently triggers a collective transformation of the entire particle. This is consistent with the abrupt transitions observed in magnetometry of single-crystalline particles. The initial nucleation events remain confined to a small surface region and are therefore not resolved in the corresponding $M(T)$-curves.

\section{Single particle XRD}

\begin{figure}
	\centering
		\includegraphics[scale=.50]{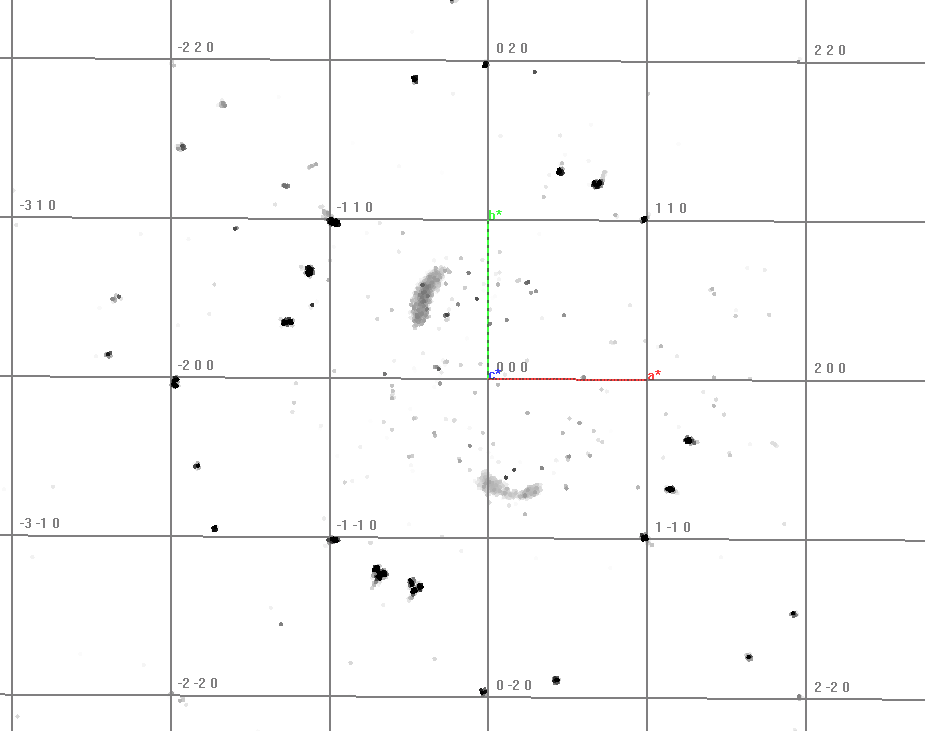}
    \caption{Section of reciprocal space (H–K–0 plane) for a representative particle ($\varnothing \approx 20\,\mu$m). Distinct reflections can be indexed by three slightly misoriented crystallites with similar lattice parameters ($a \approx 2.97$~\AA). One dominant orientation and two rotated variants are observed, indicating the presence of low-angle grain boundaries.}
    \label{fig:RSM_Plane_20mum}
\end{figure}

Figure \ref{fig:RSM_Plane_20mum} shows a section of reciprocal space (H–K–0 plane) for a representative small ($\varnothing$ = 20\,µm) particle. Despite a low signal-to-noise ratio due to significant background contributions, distinct reflections can be identified and indexed. 

The reflections can be described by three slightly misoriented crystallites with similar lattice parameters ($a \approx 2.97$~\AA, $90^\circ$). One dominant orientation is observed, accompanied by two additional sets of reflections that are slightly rotated with respect to it. This indicates the presence of low-angle grain boundaries, i.e., an oligocrystalline rather than a single-crystalline particle. 
Due to the small misorientation between crystallites, these low-angle grain boundaries are expected to have only a minor impact on the martensitic transformation compared to high-angle grain boundaries.

\begin{figure}
	\centering
		\includegraphics[scale=.50]{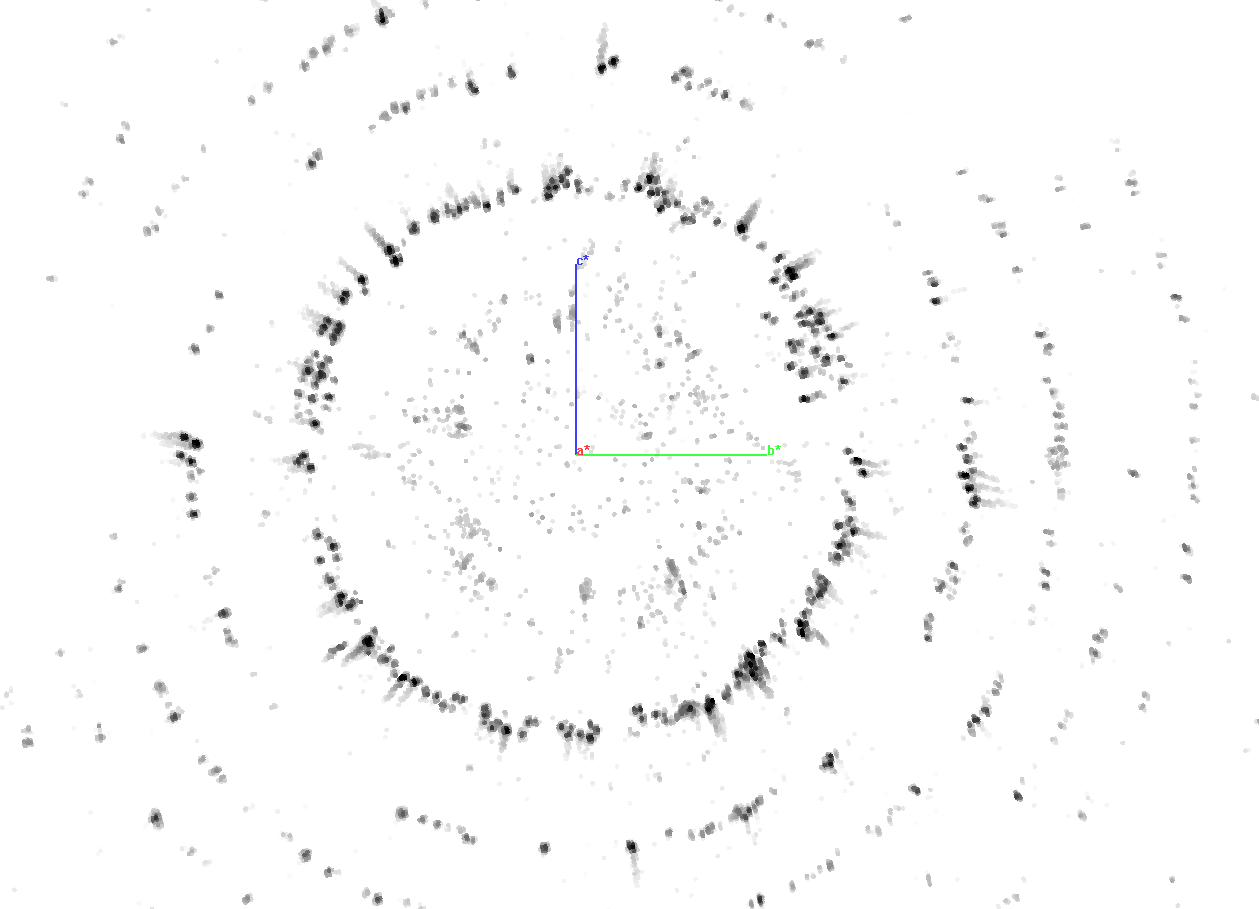}
    \caption{Section of reciprocal space (H–K–0 plane) for a larger particle ($\varnothing \approx 105\,\mu$m). The diffraction pattern shows a high density of reflections forming incomplete Debye-Scherrer rings. No unique unit cell can be assigned, consistent with a polycrystalline microstructure.}
    \label{fig:RSM_plane_105mum}
\end{figure}

In contrast, a larger particle ($\varnothing$ = 105\,µm) in Figure \ref{fig:RSM_plane_105mum} exhibits a large number of reflections, resulting in a pattern resembling incomplete Debye-Scherrer rings. In this case, no unique unit cell can be assigned, consistent with a polycrystalline structure.

\medskip

%